\documentclass[12pt]{JHEP3}
\usepackage{amsmath,amssymb,bbm}

\def\de{\partial}

\def\2{\frac12}
\def\4{\frac14}
\def\ie{{\it i.e.}}
\def\a{\alpha}
\def\b{\beta}
\def\g{\gamma}

\def\d{\delta}
\def\e{\epsilon}

\def\l{\lambda}
\def\L{\Lambda}
\def\m{\mu}
\def\n{\nu}

\def\r{\rho}
\def\s{\sigma}

\def\t{\tau}
\def\f{\phi}

\def\de{\partial}

\def\tb{\tilde{B}}
\def\tc{\tilde{C}}

\def\be{\begin{equation}}
\def\ee{\end{equation}}
\def\bea{\begin{eqnarray}}
\def\eea{\end{eqnarray}}
\author{Eric A. Bergshoeff, Mees de Roo, Sven F. Kerstan\\Centre for
Theoretical Physics, University of
Groningen, Nijenborgh 4, 9747 AG Groningen, The Netherlands\\
\email{E.A.Bergshoeff, M.de.Roo, S.Kerstan@phys.rug.nl}}
\author{Fabio Riccioni
\\ DAMTP, Centre for Mathematical Sciences,
University of Cambridge,  Wilberforce Road, Cambridge CB3 0WA, UK\\
\email{F.Riccioni@damtp.cam.ac.uk}}
\abstract{
We show in the $SU(1,1)$-covariant formulation that
IIB supergravity allows the introduction of
a doublet and a quadruplet of ten-form potentials. The
Ramond-Ramond ten-form potential which is associated with the SO(32) Type I
superstring is in the quadruplet.
Our results are consistent with
a recently proposed $E_{11}$ symmetry
underlying string theory.

For the reader's convenience we present the full
supersymmetry and gauge transformations of {\it all} fields
both in the  manifestly $SU(1,1)$ covariant
Einstein frame and in the  real $U(1)$ gauge fixed string frame.
}
\preprint{DAMTP-2005-47 \\ UG-05-04}
\title{IIB Supergravity Revisited}

\begin{document}

\section{Introduction\label{Intro}}
IIB supergravity~\cite{SW, schwarz, HW} is the low energy
effective action of type-IIB superstring theory. Its scalar sector
describes the coset manifold $SL(2,\mathbb{R})/SO(2)\simeq
SU(1,1)/U(1)$, whose isometry $SL(2,\mathbb{R})$ is a symmetry of the
low energy theory. Since the isometry acts non-trivially on the
dilaton, the full perturbative string theory does not preserve the
symmetry, but the conjecture is that non-perturbatively an
$SL(2,\mathbb{Z})$ subgroup of the full symmetry group of the low
energy action survives~\cite{ht}.

The particular feature of type-IIB string theory with respect to the
other theories of closed oriented strings is that it is symmetric
under the orientation reversal of the fundamental string.
Ten-dimensional type-I string theory is obtained from type-IIB
through an orientifold projection~\cite{augusto} that gauges this
symmetry, and tadpole cancellation requires the introduction of an
open sector, corresponding to D9-branes. The standard supersymmetric
projection gives rise to the type-I superstring, with gauge group
$SO(32)$~\cite{gs}, while a non-supersymmetric, anomaly-free
projection gives rise to a model with gauge group $USp(32)$
\cite{sugimoto}, in which supersymmetry is realized on the bulk and
spontaneously broken on the branes~\cite{bsb}.

In the low-energy effective action, the closed sector of type-I
strings is obtained by performing a consistent $\mathbb{Z}_2$ truncation of the
IIB supergravity, while the open sector corresponds to the
first order in the low-energy expansion of the D9-brane action in a
type-I background. In~\cite{BdRJO} it was shown that the $\mathbb{Z}_2$
symmetry responsible for this truncation can be performed in two
ways, and in a flat background, with all bulk fields put to zero,
the D9-brane action reduces in one case to the Volkov-Akulov
action~\cite{va}, and in the other case to a constant. In~\cite{fr}
these results were extended to a generic background, showing that
also in the curved case there are two possibilities of performing
the truncation. In one case one gets a dilaton tadpole and a RR
tadpole plus goldstino couplings, which is basically the one-brane
equivalent of the Sugimoto model, while in the other case the
goldstino couplings vanish and one is left with a dilaton and a RR
tadpole, which is the one-brane equivalent of the supersymmetric
model. In order to truncate the theory in the brane sector,
the ``democratic formulation'' of IIB
supergravity was derived \cite{BdRJO,Bergshoeff:2001pv}.
This amounts to an extension of the
supersymmetry algebra, so that both the RR fields and their magnetic
duals appear on the same footing. The closure of the algebra then
requires the field strengths of these fields to be related by
duality conditions. The result is that, together with the RR forms
$C^{(2n)}$, $n=0,...,4$ associated with D-branes of non-vanishing
codimension, the algebra naturally includes a RR ten-form $C^{(10)}$,
with respect to which the spacetime-filling D9-branes are
electrically charged. This field does not have any field
strength, and correspondingly an object charged with respect to it
can be consistently included in the theory only when one performs a
type-I truncation, so that the resulting overall RR charge vanishes.
The analysis of~\cite{BdRJO} also showed that an additional ten-form
$B^{(10)}$ can be introduced in the algebra, and this form survives
a different $\mathbb{Z}_2$ truncation, projecting out all the RR-fields. In
the string frame, the tension of a spacetime-filling brane
electrically charged with respect to $B^{(10)}$ would scale like
$g_S^{-2}$, instead of $g_S^{-4}$, thus implying that the brane
action for this object can not be obtained performing an $S$-duality
transformation on the D9-brane effective action~\cite{fabio}. We are
therefore facing a problem, since two ten-forms are known in IIB
supergravity, but they do not form a doublet with respect to
$SL(2,\mathbb{R})$.

In this paper we will clarify this issue.
We want to obtain all the possible independent ten-forms that can be
added to 10-dimensional IIB supergravity, with their assignment to
representations of $SL(2, \mathbb{R})$. In order to perform this
analysis, we express the theory in a ``$SU(1,1)$-democratic
formulation'', in which all the forms, not only the RR ones, and
their magnetic duals are described in a $SU(1,1)$-covariant way. We
use the notation of~\cite{SW, schwarz}, so that the scalars
parametrize the coset $SU(1,1)/U(1)$, while the two two-forms, as well
as their duals, form a doublet of $SU(1,1)$. The eight-forms, dual
to the scalars, transform as a triplet of $SU(1,1)$, with the field
strengths satisfying an $SU(1,1)$ invariant constraint
\cite{mo,DLT}. Eventually, we find that the algebra includes a
doublet and a quadruplet of ten-forms \footnote{Gauge fields of
maximal rank have been explored in the literature \cite{gates}.},
and the dilaton dependence of the supersymmetry transformation of
these objects shows that the RR ten-form belongs to the quadruplet.
We claim that no other independent ten-forms can be added to the
algebra. In summary, we find  the following bosonic field content:
\begin{equation}\label{extended}
e_{\m}^a, V^\a_+, V^\a_-, A^\a_{(2)}, A_{(4)}, A^\a_{(6)}, A^{(\a\b)}_{(8)}, A^\a_{(10)}, A^{(\a\b\g)}_{(10)} ,
\end{equation}
where $e_\mu^a$ is the zehnbein, $(V_+^\alpha, V_-^\alpha)$ parametrizes the
$SU(1,1)/U(1)$ coset, $\alpha=1,2$ is an $SU(1,1)$ index and the subindex
$(n)$ indicates the rank of the potential.

This paper will be devoted to the construction
and the properties of the extended IIB supergravity theory
(\ref{extended}). Clearly
the properties of the dual forms and ten-forms have implications for
the structure of the brane spectrum, dualities, etc. These aspects of this
work will be addressed in a forthcoming paper \cite{BDKOR}.

The structure of the paper is as follows. The main result, the
supersymmetry transformation rules and algebra of the extended
IIB-supergravity theory in the $SU(1,1)/U(1)$ formulation, are given
in section \ref{results}. In section \ref{SF} these results are
rewritten in a $U(1)$ gauge in the Einstein frame and in the string
frame. In this section we also recover the Ramond-Ramond
``harmonica'' of \cite{BdRJO} and then extend it to the
Neveu-Schwarz forms. We also list the action of $S$-duality on all
form fields. The preceding sections lead up to these results and
sketch the derivation. In section \ref{su11} we review the
$SU(1,1)$-covariant notation of \cite{SW, schwarz}. In section
\ref{extensions} we introduce in the algebra the six- and the
eight-forms dual to the two-forms and the scalars respectively.
Section \ref{tenforms} contains the analysis of the ten-forms. We
finally conclude with a summary of our results and a discussion.
Some basic formulas and truncations to $N=1$ supergravity can be
found in the Appendices.

\section{The $SU(1,1)$-covariant formulation}\label{su11}

In this section we review the notation and the results
of \cite{SW,schwarz}.\\
The theory contains the graviton, two scalars, two two-forms and a
self-dual four-form in the bosonic sector, together with a complex
left-handed gravitino and a complex right-handed spinor in the
fermionic sector. We will use the mostly-minus spacetime signature
convention throughout the paper. The two scalars parametrize the
coset $SU(1,1)/U(1)$, that can be described in terms of the
$SU(1,1)$ matrix ($\alpha,\beta =1,2$)
  \begin{equation}
  U = ( \ V_-^\a \ \ V_+^\a \ ) \quad ,
  \end{equation}
satisfying the constraint
  \begin{equation}
  V_-^\a V_+^\b - V_+^\a V_-^\b = \e^{\a\b} \quad , \label{VVe}
  \end{equation}
with $(V^1_- )^* = V^2_+$, where $\a=1,2$ is an $SU(1,1)$ index and
$+$ and $-$ denote the $U(1)$ charge, and $\e^{12}= \e_{12}=1$. From
the left-invariant 1-form
  \begin{equation}
  U^{-1} \de_\m U = \left(
  \begin{array}{cc} -i Q_\m & P_\m
  \\ P_\m^* & iQ_\m \end{array} \right) \label{scalarmatrix}
  \end{equation} one reads off the $U(1)$-covariant quantity
  \begin{equation}
  P_\m = -\e_{\a\b}  V_+^\a
  \de_\m V_+^\b \quad ,
  \end{equation}
that has charge 2, and the $U(1)$ connection
  \begin{equation}
  Q_\m = -i \e_{\a\b}
  V_-^\a  \de_\m V_+^\b \quad .
  \end{equation}
Note that
\begin{eqnarray}
    P_\m V^\a_- &=&   D_\m V^\a_+ \quad,\\
    P^*_\m V^\a_+ &=& D_\m V^\a_-\quad ,
\end{eqnarray}
where the derivative $D$ is covariant with respect to $U(1)$.
The two-forms are collected in an $SU(1,1)$ doublet $A^\a_{\m\n}$
satisfying the constraint
  \begin{equation}
  (A^1_{\m\n} )^* = A^2_{\m\n} \quad .
  \end{equation}
The corresponding field strengths
  \begin{equation}
  F^\a_{\m\n\r} = 3 \de_{[\m} A^\a_{\n\r ]}
  \end{equation}
are invariant with respect to the gauge transformations
  \begin{equation}
  \d A^\a_{\m\n} = 2 \de_{[ \m} \L^\a_{\n ]}\quad .
  \end{equation}
The four-form is invariant under $SU(1,1)$, and varies as
  \begin{equation}
  \d A_{\m\n\r\s} = 4 \de_{[\m}\L_{\n\r\s]}-\frac{i}{4} \e_{\a\b}
  \L^{\a}_{[\m} F^\b_{\n\r\s ]} \label{4formgauge}
  \end{equation}
under four-form and two-form gauge transformations, so that the
gauge-invariant five-form field-strength is
  \begin{equation}
  F_{\m\n\r\s\t} =5 \de_{[\m} A_{\n\r\s\t ]}+
  \frac{5i}{8} \e_{\a\b} A^\a_{[\m\n} F^\b_{\r\s\t ]} \quad
  .\label{5formfieldstrength}
  \end{equation}
This five-form satisfies the self-duality condition
  \begin{equation} F^{\m_1 ... \m_5 } = \frac{1}{5 !}
  \e^{\m_1 ...\m_5 \n_1 ... \n_5 }F_{\n_1 ...\n_5} \quad . \label{sd5}
  \end{equation}
It is convenient to define the complex three-form
  \begin{equation}
  G_{\m\n\r}^{} = - \e_{\a\b} V^\a_+ F^\b_{\m\n\r} \quad ,
  \end{equation}
that is an $SU(1,1)$ singlet with $U(1)$ charge 1. Finally the
gravitino $\psi_\m$ is complex left-handed with $U(1)$ charge $1/2$,
while the spinor $\l$ is complex right-handed with $U(1)$ charge
$3/2$.

In \cite{schwarz} the field equations for this model
were derived by requiring the closure of the
supersymmetry algebra. All these equations can be derived from a
lagrangian, imposing eq. (\ref{sd5}) only after varying
\cite{BHO}\footnote{A
lagrangian formulation for self dual forms has been developed
in~\cite{pst}, and then applied in~\cite{DLT} to the ten-dimensional
IIB supergravity. It corresponds to the introduction of an
additional scalar auxiliary field, and the self-duality condition
results from the gauge fixing (that can not be imposed directly on
the action) of additional local symmetries.}. It is interesting to
study in detail the kinetic term for the scalar fields,
  \begin{equation}
  {\cal L}_{scalar} =\frac{e}{2} P^*_\m P^\m  \quad .
  \end{equation}
The complex variable
\begin{equation}
  z=\frac{V_-^2}{V^1_-} \label{defz}
\end{equation}
is invariant
under local $U(1)$ transformations, and so it is a good coordinate
for the scalar manifold. Under the $SU(1,1)$ transformation
  \begin{equation}
  \left( \begin{array}{c} V^1_- \\ V^2_- \end{array} \right)
  \rightarrow \left( \begin{array}{cc} \a & \b \\
  \bar{\b} & \bar{\a} \end{array} \right) \left(
  \begin{array}{c} V^1_- \\ V^2_- \end{array} \right)
  \quad ,\label{su11transf}
  \end{equation}
that is an isometry of the scalar manifold, $z$ transforms as
  \begin{equation}
  z \rightarrow    \frac{\bar{\a} z +\bar{\b}}{\b z +\a } \quad .
  \end{equation}
The variable $z$ parametrizes the unit disc, $\vert z \vert < 1$,
and the kinetic term assumes the form
  \begin{equation}
  {\cal L}_{scalar} =-\frac{e}{2}
  \frac{\de_\m z \de^\m \bar{z}}{(1 - z\bar{z})^2} \quad.
  \end{equation}
The further change of variables
  \begin{equation}
  z = \frac{1+i\t}{1-i\t}
\label{deftau}
  \end{equation}
maps the disc in the complex upper-half plane, ${\rm Im} \t > 0$,
and in terms of $\t$ the transformations (\ref{su11transf}) become
  \begin{equation}
  \t \rightarrow \frac{a \t+ b}{c\t +d } \quad ,\label{sl2r}
  \end{equation}
where
  \begin{equation}
  \left( \begin{array}{cc} a & b \\ c & d \end{array}
  \right) \in SL(2,\mathbb{R}) \quad ,
  \end{equation}
while the scalar lagrangian takes the form
  \begin{equation}
  {\cal L}_{scalar} =-\frac{e}{8}
  \frac{\de_\m \t \de^\m \bar{\t}}{({\rm Im} \t)^2} \quad .
  \end{equation}
Expressing $\t$ in terms of the RR scalar and the dilaton,
  \begin{equation}
  \t = \ell+ i e^{-\phi} \label{deftlf}
  \end{equation}
and performing the Weyl rescaling $g_{(E)\m\n} \rightarrow e^{-\phi /2 }
g_{(S)\m\n}$ one ends up with the standard form of the kinetic term of
the scalars in IIB supergravity in the string frame.

The supersymmetry transformations that leave the field equations of
\cite{schwarz} invariant are
  \begin{eqnarray}
  & & \d e_\m{}^a = i \bar{\e} \g^a \psi_\m
      + i \bar{\e}_C \g^a \psi_{\m C}  \quad ,
\nonumber\\
  & & \d \psi_\m = D_\m \e +\tfrac{i}{480} F_{\m\n_1 ...\n_4 } \g^{\n_1 ...\n_4 } \e
  +\tfrac{1}{96} G^{\n\r\s} \g_{\m\n\r\s} \e_C
  -\tfrac{3}{32}
  G_{\m\n\r}^{} \g^{\n\r} \e_C \quad ,
\nonumber \\
  & & \d A^\a_{\m\n} = V^\a_- \ \bar{\e} \g_{\m\n} \l
                      +V^\a_+ \ \bar{\e}_C \g_{\m\n} \l_C
  +4i V^\a_- \ \bar{\e}_C \g_{[\m} \psi_{\n ]}
  +4i V^\a_+ \ \bar{\e} \g_{[\m} \psi_{\n ]C}
  \quad ,
\nonumber \\
  & & \d A_{\m\n\r\s} =\bar{\e} \g_{[ \m\n\r} \psi_{\s ]}
                      -\bar{\e}_C \g_{[ \m\n\r} \psi_{\s ]C}
  -\tfrac{3i}{8} \e_{\a\b} A^\a_{[\m\n} \d A^\b_{\r \s ]} \quad ,
\nonumber \\
  & & \d \l = i P_\m \g^\m \e_C -\tfrac{i}{24}
  G_{\m\n\r}^{} \g^{\m\n\r} \e \quad ,
\nonumber \\
  & & \d V^\a_+ = V^\a_- \ \bar{\e}_C \l  \quad ,
\nonumber \\
  & & \d V^\a_- = V^\a_+ \ \bar{\e} \l_C   \quad . \label{susyschwarz}
  \end{eqnarray}
where we denote with $\Psi_C$ the complex (Majorana) conjugate of
$\Psi$. The commutator $[\d_1 , \d_2 ]$ of two supersymmetry
transformations of (\ref{susyschwarz}) closes on all the local
symmetries of the theory, provided one uses the fermionic field
equations and the self-duality condition of eq. (\ref{sd5}). To
lowest order in the fermions, the parameters of the resulting
general coordinate transformation, four-form gauge transformation and
two-form gauge transformation are\footnote{We only present the
parameters of translations and the two- and four-form gaugetransformations.
The parameters of other local symmetries,
namely supersymmetry, local Lorentz and
local $U(1)$ are not used in the analysis of the
next sections, and are given in \cite{SW, schwarz}.}
  \begin{eqnarray}
  & & \xi^\m = i \ \bar{\e}_2 \g^\m \e_1 +
  i \ \bar{\e}_{2C} \g^\m \e_{1C} \quad , \nonumber \\
  & & \L^\a_\m = A_{\m\n}^\a \xi^\n - 2 i [ V^\a_+ \ \bar{\e}_2
  \g_\m \e_{1C} + V^\a_- \ \bar{\e}_{2C}
  \g_\m \e_{1} ] \quad , \nonumber \\
  & & \L_{\m\n\r} = A_{\m\n\r\s} \xi^\s -\frac{1}{4} [ \bar{\e}_2
  \g_{\m\n\r} \e_1 - \bar{\e}_{2C} \g_{\m\n\r} \e_{1C} ] \nonumber
  \\
  & & \quad \quad - \tfrac{3}{8} \e_{\a\b} A^\a_{[\m\n} \big( V^\b_+ \ \bar{\e}_2
  \g_{\r]} \e_{1C} + V^\b_- \ \bar{\e}_{2C}
  \g_{\r]} \e_{1} \big) \quad . \label{parameters}
  \end{eqnarray}
In the next section we will extend the algebra in order to include
the magnetic duals of the scalars and of the two-form, in such a way
that the supersymmetry algebra still closes, once the proper duality
relations are used. Once we obtain the supersymmetry transformation
of the six- and the eight-forms that are compatible with the algebra
obtained from eq. (\ref{susyschwarz}), we will include in Section
\ref{tenforms}
all the possible independent ten-forms that this algebra allows.

\section{Six-forms and eight-forms} \label{extensions}
In this section we show how the algebra of eq. (\ref{susyschwarz})
is extended introducing the forms magnetically dual to the scalars
and the two-forms. As anticipated, closure of the supersymmetry
algebra requires the field strengths of these forms to be related to
$P_\m$ and the field strengths of the two-forms by suitable duality
relations. Generalizing what happens for the four-form (see eqs.
(\ref{4formgauge}) and (\ref{5formfieldstrength})), we will see that
the gauge transformations of these fields involve the gauge
parameters of all the lower rank forms, and the gauge invariant
field strengths will therefore contain lower rank forms as well.
After introducing our Ansatz for these field strengths and gauge
transformations, the supersymmetry transformations of these fields
will then be determined requiring the closure of the supersymmetry
algebra. As in the previous section, we will not consider terms
higher than quadratic in the fermi fields.

\subsection{Six-forms\label{sixforms}}
We want to obtain the gauge and supersymmetry transformations for
the doublet of six-forms $A^\a_{\m_1 \dots \m_6}$, which are the
magnetic duals of the two-forms and thus satisfy the reality
condition
  \begin{equation}
  (A^1 )^*_{\m_1\dots \m_6} = A^2_{\m_1 \dots \m_6} \quad .
  \label{real6form}
  \end{equation}
Generalizing what one obtains for the four-form, we expect the
supersymmetry transformation of the six-forms to contain terms
involving only spinors and terms containing forms of lower rank. The
condition of eq. (\ref{real6form}), as well as the requirement that
all the terms must have vanishing local $U(1)$ charge, fixes the
most general transformation of the  doublet to be
  \begin{eqnarray}
  \d A_{\m_1 \cdots \m_6}^{\a} &=& a \
  V^\a_- \ \bar{\e} \g_{\m_1 ...\m_6} \l + a^* \ V^\a_+ \ \bar{\e}_C
  \g_{\m_1 ...\m_6} \l_C
  \nonumber \\
  &+&  b \ V^\a_- \ \bar{\e}_C \g_{[\m_1 ... \m_5 }
  \psi_{\m_6]} - b^* \ V^\a_+ \ \bar{\e} \g_{[\m_1 ... \m_5 } \psi_{\m_6] C}
  \nonumber\\
  &+& c \ A_{[\m_1\cdots\m_4 }\d
  A^{\a}_{\m_5\m_6 ]}
  \nonumber\\
  &+& d \ \d A_{[\m_1\cdots\m_4 }
  A^{\a}_{\m_5\m_6 ]}
  \nonumber\\
  &+& i e \ \e_{\b\g}\d A^\b_{[\m_1\m_2} A^\g_{\m_3 \m_4} A^\a_{\m_5 \m_6]}
  \quad .  \label{susy6form}
  \end{eqnarray}
We want to consider the commutator $[ \delta_1 , \delta_2 ]$ of two
such transformations, to lowest order in the fermi fields.

We first take into account the terms involving the spinors, {\ie},
the first two lines in eq. (\ref{susy6form}). Those terms produce
the gauge transformation for the six-forms
  \begin{eqnarray}
  \d A^\a_{\m_1 \dots \m_6} &=& 6 \de_{[\m_1 } \L^\a_{\m_2 \dots
  \m_6]} \nonumber \\
  &=& -12 i\de_{[\m_1} ( a \ V^\a_+ \ \bar{\e}_2 \g_{\m_2 \dots \m_6 ]}
  \e_{1C}
  + a^* \  V^\a_- \ \bar{\e}_{2C} \g_{\m_2 \dots \m_6 ]} \e_{1} )
  \label{lambda5}
  \end{eqnarray}
if the constraint
  \begin{equation}
  12 i a^* = b
  \end{equation}
is imposed, while the other terms that are produced are
  \begin{eqnarray}
  & &   20 i a F^\a_{[\m_1 \m_2 \m_3} (
  \bar{\e}_{2C} \g_{\m_4 \m_5 \m_6] } \e_{1C} - \bar{\e}_{2}
  \g_{\m_4
  \m_5 \m_6 ]} \e_{1})\nonumber \\
  & & - \tfrac{1}{6}a  \e_{\m_1 ...\m_6 \sigma \m\n\r}
  S^{\a\b} \e_{\b\g} F^{\g ; \m\n\r} \xi^\sigma \quad
  ,\label{atermcomm}
  \end{eqnarray}
  where we have defined
  \begin{equation}
  S^{\a\b}= V^\a_- V^\b_+ + V^\a_+ V^\b_-
  \end{equation}
and we have assumed that $a$ is imaginary.
Note that $S^{\a\b}$ satisfies
\begin{equation}
   S^{\a\b}\e_{\b\g} S^{\g\d} \e_{\d\e} = \d^\a_\e\quad.
\end{equation}
Observe that there
are no terms involving the five-form field strength. Without loss of
generality, we fix
  \begin{equation} a = i
  \end{equation}
from now on. In order for the last term in (\ref{atermcomm}) to
produce a general coordinate transformation with the right
coefficient as dictated by eq. (\ref{parameters}), we impose the
duality relation \footnote{Note that this duality relation induces field
equations for the potentials.}
  \begin{equation}
  F^\a_{\m_1 ... \m_7 } = -\tfrac{i}{3!}
  \e_{\m_1 ... \m_7 \m\n\r} S^{\a\b} \e_{\b\g} F^{\g ; \m\n\r} \quad
  ,
  \end{equation}
where $F^\a_{\m_1 \dots \m_7} = 7 \de_{[\m_1}A^\a_{\m_2 \dots \m_7]}
+ \dots $ are the field strengths of the six-forms, and the dots
stand for terms involving lower rank forms that we will determine in
the following. Note that the second term of eq. (\ref{atermcomm})
contains, together with a general coordinate transformation, a gauge
transformation of parameter
\begin{equation}
  \L'{}^\a_{\m_1 \dots \m_5} = A_{\m_1 \dots \m_5 \s}^\a \xi^\s \quad
  . \label{lambda5prime}
\end{equation}
The $SU(1,1)$-invariant quantities
\begin{eqnarray}
  G^{}_{\m_1 ... \m_7} = -\e_{\a\b} V^\a_+
  F^\b_{\m_1 ... \m_7} \,,\quad
  G^{*}_{\m_1 ... \m_7} =
  \e_{\a\b} V^\a_- F^\b_{\m_1 ... \m_7} \quad ,
\end{eqnarray}
 which have $U(1)$ charge $+1$ and $-1$ respectively, satisfy
\begin{eqnarray} G^{(7)}_{\m_1 ... \m_7} = \tfrac{i}{3!}
  \e_{\m_1 ... \m_7 \m\n\r} G^{\m\n\r} \,,\quad
 G^{*}_{\m_1 ...
  \m_7} = -\tfrac{i}{3!} \e_{\m_1 ... \m_7 \m\n\r} G^{*\,\m\n\r} \quad  .
\end{eqnarray}

In order to proceed further,  in analogy with eq.
(\ref{5formfieldstrength}) we make the following Ansatz for the
seven-form field strengths:
\begin{equation}
   F^\a_{\m_1 \dots \m_7} = 7 \de_{[\m_1} A^\a_{\m_2 \dots \m_7 ]} +
  \a A^\a_{[\m_1 \m_2 } F_{\m_3 \dots \m_7 ]} +
  \b F^\a_{[\m_1 \dots \m_3}A_{\m_4 \dots \m_7]}
  \quad . \label{7formfieldstrength}
\end{equation}
For these forms to be gauge invariant, the  must transform
non-trivially with respect to the two-form and four-form gauge
transformations. The result is
\begin{equation}
  \d A^\a_{\m_1 \dots \m_6} = -\tfrac{2}{7}\a
  \L^\a_{[\m_1} F_{\m_2 \dots \m_6]} +
  \tfrac{4}{7}\b F^\a_{[\m_1 \dots \m_3}\L_{\m_4 \dots \m_6 ]} \quad
  ,\label{extragauge6}
  \end{equation}
and gauge invariance requires
  \begin{equation} \b = -\frac{10}{3} \a \quad .
  \end{equation}

Now we come back to the commutator. The terms that are left are the
ones coming from the last three lines in eq. (\ref{susy6form}),
together with the first line in eq. (\ref{atermcomm}) and the terms
coming from (\ref{7formfieldstrength}) in the second line of eq.
(\ref{atermcomm}). All these terms have to produce gauge
transformations according to (\ref{extragauge6}), with parameters
given from eqs. (\ref{parameters}), possibly together with
additional  gauge transformations. The end result is that one
produces the additional  gauge transformations
  \begin{eqnarray}
  \L'{}'{}^\a_{\m_1 \dots \m_5} &=& -\tfrac{2i}{3} c \ A_{[\m_1 \dots \m_4}
  (
  V^\a_+ \ \bar{\e}_2
  \g_{\m_5 ]} \e_{1C} + V^\a_- \ \bar{\e}_{2C}
  \g_{\m_5 ]} \e_{1}  )\nonumber \\
  & - & \tfrac{1}{6} d \ A^\a_{[\m_1 \m_2} ( \bar{\e}_2
  \g_{\m_3 \dots \m_5 ]} \e_1 - \bar{\e}_{2C} \g_{\m_3 \dots \m_5 ]} \e_{1C}
  )
  \quad , \label{lambda5doubleprime}
  \end{eqnarray}
while all the coefficients are uniquely determined to be
  \begin{equation} c = 40  \quad , \qquad d = -20 \quad , \qquad
  e = \frac{15}{2}  \quad , \qquad
  \a=28 \quad .
  \end{equation}
Summarizing, we get that the supersymmetry transformations of the
six-forms are
  \begin{eqnarray}
  \d A_{\m_1 \cdots \m_6}^{\a} &=& i \
  V^\a_- \ \bar{\e} \g_{\m_1 ...\m_6} \l -i \ V^\a_+ \ \bar{\e}_C
  \g_{\m_1 ...\m_6} \l_C
  \nonumber \\
  &+&  12 \ V^\a_- \ \bar{\e}_C \g_{[\m_1 ... \m_5 }
  \psi_{\m_6]} - 12 \ V^\a_+ \ \bar{\e} \g_{[\m_1 ... \m_5 } \psi_{\m_6] C}
  \nonumber\\
  &+& 40 \ A_{[\m_1\cdots\m_4 }\d
  A^{\a}_{\m_5\m_6 ]}
  \nonumber\\
  &-& 20 \ \d A_{[\m_1\cdots\m_4 }
  A^{\a}_{\m_5\m_6 ]}
  \nonumber\\
  &+&  \tfrac{15i}{2} \e_{\b\g}\d A^\b_{[\m_1\m_2} A^\g_{\m_3 \m_4} A^\a_{\m_5 \m_6]}
  \quad .  \label{susy6form2}
  \end{eqnarray}
The doublet of seven-form field strengths is
  \begin{equation} F^\a_{\m_1 \dots \m_7} = 7 \de_{[\m_1} A^\a_{\m_2 \dots \m_7]} +
  28 A^\a_{[\m_1 \m_2} F_{\m_3 \dots \m_7 ]} -\tfrac{280}{3} F^\a_{[\m_1 \dots \m_3}
  A_{\m_4 \dots \m_7 ]} \quad . \label{7formfieldstrength2}
  \end{equation}
This is gauge invariant with respect to the transformations of the
two-forms, the four-form and the six-forms, where
  \begin{equation} \d A^\a_{\m_1 \dots \m_6} = 6 \de_{[\m_1} \L^\a_{\m_2 \dots \m_6
  ]}- 8
  \L^\a_{[\m_1} F_{\m_2 \dots \m_6]} - \tfrac{160}{3} F^\a_{[\m_1 \dots \m_3}\L_{\m_4 \dots \m_6 ]} \quad
  .\label{extragauge62}
  \end{equation}
Moreover, the six-form gauge transformation parameter resulting from
the commutator of two supersymmetry transformations is
  \begin{eqnarray}
  \L^\a_{\m_1 \dots \m_5} &=&
  A^\a_{\m_1 \dots \m_5 \s}\xi^\s +
 2(V^\a_+ \ \bar{\e}_2 \g_{\m_1 \dots \m_5} \e_1
  - V^\a_- \ \bar{\e}_{2C} \g_{\m_1 \dots \m_5} \e_{1C} )  \nonumber \\
  &-&\tfrac{80i}{3} A_{[\m_1 \dots \m_4} (
  V^\a_+ \ \bar{\e}_2
  \g_{\m_5 ]} \e_{1C} + V^\a_- \ \bar{\e}_{2C}
  \g_{\m_5 ]} \e_{1}  )\nonumber \\
  & + & \tfrac{10}{3}  A^\a_{[\m_1 \m_2} ( \bar{\e}_2
  \g_{\m_3 \dots \m_5 ]} \e_1 - \bar{\e}_{2C} \g_{\m_3 \dots \m_5 ]} \e_{1C}
  ) \quad ,\label{5formparameter}
  \end{eqnarray}
as results from eqs. (\ref{lambda5}), (\ref{lambda5prime}) and
(\ref{lambda5doubleprime}). Finally, a comment is in order. At
first sight, the Ansatz we made for the field strengths in eq.
(\ref{7formfieldstrength}) does not seem to be the most general one,
since one could in principle include a term of the form $i \e_{\b\g}
A^a_{[\m_1 \m_2} A^\b_{\m_3 \m_4} F^\g_{\m_5\dots \m_7]}$. The
reason why we did not include it is that one can always reabsorb
such a term by performing a redefinition of the six-forms of the type
$A^\a_{\m_1 \dots \m_6} \rightarrow A^\a_{\m_1 \dots \m_6} + \g
A^\a_{[\m_1 \m_2} A_{\m_3 \dots \m_6]}$, and choose $\g$ so that
this term vanishes. This freedom will be used to constrain the form
of the field strengths of the eight-forms as well, as we will see in the
next subsection.

\subsection{Eight-forms\label{eightforms}}
The eight-forms are the magnetic duals of the scalars. As we
reviewed in Section \ref{su11}, the scalars are described in terms
of the left-invariant 1-form of eq. (\ref{scalarmatrix}),
transforming in the adjoint of $SU(1,1)$, and propagating two real
degrees of freedom because of local $U(1)$ invariance. One therefore
expects a triplet of eight-forms (as observed in \cite{mo,DLT}) \footnote{A
similar observation was made for the curvatures in \cite{Juliaetal}.},
that we denote by $A^{\a\b}_{\m_1 \dots \m_8}$, symmetric under $\a
\leftrightarrow \b$, and satisfying the reality condition
  \begin{equation}
  (A^{11})^*_{\m_1 \dots \m_8} = A^{22}_{\m_1 \dots \m_8} \quad ,
  \qquad (A^{12})^*_{\m_1 \dots \m_8} = A^{12}_{\m_1 \dots \m_8} \quad .
  \end{equation}
The fact that only two scalars propagate will result in a constraint
for the field strengths of these eight-forms \cite{Juliaetal,DLT}. This is
exactly what we are going to show in this subsection. Following the
same arguments as in the previous subsection, we write the most
general supersymmetry transformations for the eight-forms,
compatible with the reality condition and with $U(1)$ invariance,
consisting of terms that only involve the spinors and terms
containing the lower rank forms and their supersymmetry
transformations. The result is
  \begin{eqnarray} \d
  A^{\a\b}_{\m_1 \dots \m_8} &=&
  a \ V^\a_+ V^\b_+ \ \bar{\e} \g_{\m_1 \dots \m_8} \l_C
  + a^* \ V^\a_- V^\b_- \ \bar{\e}_C \g_{\m_1 \dots \m_8} \l \nonumber\\
  &+& b \ V^{(\a}_+ V^{\b)}_- \ \bar{\e} \g_{[\m_1 \dots \m_7} \psi_{\m_8 ]} -
  b^* \ V^{(\a}_+ V^{\b)}_- \ \bar{\e}_C \g_{[\m_1 \dots \m_7} \psi_{\m_8 ]C}
  \nonumber \\
  &+& c A^{(\a}_{[\m_1 \dots \m_6} \d A^{\b)}_{\m_7 \m_8]}
  +  d A^{(\a}_{[\m_1 \m_2} \d A^{\b)}_{\m_3 \dots \m_8]}
  + i e A^{(\a}_{[\m_1 \m_2} A^{\b)}_{\m_3 \m_4} \e_{\g\d} A^\g_{\m_5 \m_6} \d A^\d_{\m_7\m_8]}\nonumber \\
  & +& f A^{(\a}_{[\m_1 \m_2} A^{\b)}_{\m_3 \m_4} \d A_{\m_5 \dots \m_8 ]}
  +   g A_{[\m_1 \dots \m_4} A^{(\a}_{\m_5 \m_6} \d A^{\b)}_{\m_7 \m_8]}
  \quad . \label{susy8}
  \end{eqnarray}
We first consider the contributions coming from the first two lines
of eq. (\ref{susy8}), in order to get a relation between $a$ and
$b$. We obtain the  gauge transformation
  \begin{eqnarray}
  \d A^{\a\b}_{\m_1 \dots \m_8 }
  & = & 8 \de_{[\m_1} \L^{\a\b}_{\m_2 \dots \m_8 ]}\nonumber \\
  &=& - 4ia \de_{[\m_1 } \left[ S^{\a\b} ( \bar{\e}_2 \g_{\m_2
  \dots \m_8 ]} \e_1 -  \bar{\e}_{2C} \g_{\m_2
  \dots \m_8 ]} \e_{1C}) \right] \label{lambda7}
  \end{eqnarray}
together with the terms
  \begin{eqnarray}
  & & 28 ia (V^{(\a}_+ \ \bar{\e}_2 \g_{[\m_1 \dots \m_5} \e_1 -
  V^{(\a}_- \ \bar{\e}_{2C} \g_{[\m_1 \dots \m_5} \e_{1C} ) F^{\b)}_{\m_6 \dots \m_8 ]} \nonumber \\
  & & - 4 a ( V^{(\a}_+ \ \bar{\e}_2 \g_{[\m_1 } \e_1 +
  V^{(\a}_- \ \bar{\e}_{2C} \g_{[\m_1 } \e_{1C} ) F^{\b)}_{\m_2 \dots \m_8]}\nonumber \\
  & &-  a \e_{\m_1 \dots \m_8 \s\t} \xi^\s ( V^\a_+ V^\b_+ P^{* \t} - V^\a_- V^\b_- P^\t ) \quad , \label{extracomm8}
  \end{eqnarray}
provided that
  \begin{equation}
  8ia = b
  \end{equation}
and $a$ is chosen to be imaginary. Fixing, without loss of
generality,
  \begin{equation}
  a= -i \quad ,
  \end{equation}
one finds that the last term in eq. (\ref{extracomm8}) contains the
correct general coordinate transformation, plus an  gauge
transformation of parameter
  \begin{equation}
  \L'{}^{\a\b}_{\m_1 \dots \m_7} = A^{\a\b}_{\m_1 \dots \m_7 \s}
  \xi^\s \label{lambda7prime}
  \end{equation}
provided the duality relation
  \begin{equation}
  F^{\a\b}_{\m_1 \dots \m_9} = i \e_{\m_1 \dots \m_9}{}^\s [
  V^\a_+ V^\b_+ P^*_\s - V^\a_- V^\b_- P_\s ]
  \label{duality91}
  \end{equation}
holds, where $F^{\a\b}_{\m_1 \dots \m_9}= 9 \de_{[\m_1}
A^{\a\b}_{\m_2 \dots \m_9 ]}+ \dots$, and the dots stand for terms
involving lower rank forms. From the field strengths of the eight-forms,
one can define the $SU(1,1)$ invariant quantity
  \begin{equation}
  G^{}_{\m_1 \dots \m_9}=\e_{\a\g} \e_{\b\d}
  V^\a_+ V^\b_+ F^{\g\d}_{\m_1 \dots \m_9} \quad ,
  \end{equation}
with $U(1)$ charge $+2$, and its complex conjugate
  \begin{equation}
  G^{*}_{\m_1 \dots \m_9} = \e_{\a\g} \e_{\b\d} V^\a_- V^\b_-
  F^{\g\d}_{\m_1 \dots \m_9} \quad .
  \end{equation}
In terms of these objects, the duality relation of eq.
(\ref{duality91}) becomes
  \begin{equation}
  G^{}_{\m_1 \dots \m_9} = -i \e_{\m_1 \dots \m_9 \s} P^\s
  \,,\quad
  G_{\m_1 \dots \m_9}^{*} = i \e_{\m_1 \dots \m_9 \s} P^{*\s} \quad .
  \end{equation}
One can define a third nine-form,
  \begin{equation} \tilde{G}^{}_{\m_1 \dots \m_9} = \e_{\a\g}
  \e_{\b\d} V^\a_+ V^\b_- F^{\g\d}_{\m_1 \dots \m_9} \quad , \label{g9tilde}
  \end{equation}
with vanishing $U(1)$ charge, but the duality relation
(\ref{duality91}) implies that this nine-form vanishes identically
\cite{DLT}, thus determining an $SU(1,1)$ invariant constraint.
Therefore
only two eight-forms are actually
independent.

We now come to our choice for the field strengths, for which the most general
general expression is
  \begin{eqnarray}
  F^{\a\b}_{\m_1 \dots \m_9} = & & 9 \de_{[\m_1}
  A^{\a\b}_{\m_2 \dots \m_9 ]} + \a F^{(\a}_{[\m_1 \dots \m_7} A^{\b)}_{\m_8 \m_9]}
  + \b F^{(\a}_{[\m_1 \dots \m_3}
  A^{\b)}_{\m_4 \dots \m_9 ]} + \g F_{[\m_1 \dots \m_5} A^{(\a}_{\m_6 \m_7} A^{\b)}_{\m_8 \m_9 ]} \nonumber \\
  & & + i \d \e_{\g\d} A^\g_{[\m_1 \m_2} F^\d_{\m_3 \dots \m_5} A^{(\a}_{\m_6 \m_7} A^{\b)}_{\m_8 \m_9 ]}
  + \xi A_{[\m_1 \dots \m_4} F^{(\a}_{\m_5 \dots \m_7} A^{\b)}_{\m_8 \m_9]} \quad . \label{9formgeneral}
  \end{eqnarray}
The freedom of redefining the eight-form, $A_8 \rightarrow A_8 + A_6 A_2
+ A_4 A_2 A_2$, can be used to put to zero the coefficients $\xi$
and $\d$ in (\ref{9formgeneral}). It turns out that defining the
gauge transformation of the eight-forms as
  \begin{equation}
  \delta A^{\a\b}_{\m_1 \dots \m_8} = 8 \de_{[\m_1}\L^{\a\b}_{\m_2 \dots \m_8 ]} +
  \tfrac{2}{9}\a F^{(\a}_{[\m_1 \dots \m_7} \L^{\b)}_{\m_8]}  +
  \tfrac{2}{3}\b F^{(\a}_{[\m_1 \dots \m_3} \L^{\b)}_{\m_4 \dots \m_8]} \quad ,
  \end{equation}
the field strengths of eq. (\ref{9formgeneral}) are gauge invariant
if the coefficient $\g$ vanishes as well, and if the coefficients
$\a$ and $\b$ are related by
  \begin{equation}
  \b = -7 \a \quad .
  \end{equation}
To summarize, we have obtained
  \begin{eqnarray}
  F^{\a\b}_{\m_1 \dots \m_9} &=& 9 \de_{[\m_1} A^{\a\b}_{\m_2 \dots \m_9]} +
  \a F^{(\a}_{[\m_1 \dots \m_7} A^{\b)}_{\m_8 \m_9 ]} -7\a F^{(\a}_{\m_1 \dots \m_3} A^{\b)}_{\m_4 \dots \m_9 ]} \quad ,
\\
  \d A^{\a\b}_{\m_1 \dots \m_8} &=& 8 \de_{[\m_1}
  \L^{\a\b}_{\m_2 \dots \m_8]} +\tfrac{2}{9}\a F^{(\a}_{[\m_1 \dots \m_7} \L^{\b)}_{\m_8]}
  - \tfrac{14}{3}\a F^{(\a}_{[ \m_1 \dots \m_3} \L^{\b)}_{\m_4 \dots \m_8 ]} \quad .
  \label{gauge8form}
  \end{eqnarray}

We now consider the terms in the commutator coming from the last two
lines of eq. (\ref{susy8}), as well as the first two terms in eq.
(\ref{extracomm8}) and the part of the third containing lower rank
forms. All these terms have to produce the gauge transformations of
eq. (\ref{gauge8form}) with the parameters given in eqs.
(\ref{parameters}) and (\ref{5formparameter}), plus possibly a
 gauge transformation. The end result is that one produces the
additional  gauge transformation
  \begin{eqnarray}
  \L^{''\a\b}_{\m_1 \dots \m_7}& =&  -4i c \
  \left[ A^{(\a}_{[\m_1 \dots \m_6} (V^{\b)}_+ \ \bar{\e}_2 \g_{\m_7 ]} \e_1
  +  V^{\b)}_- \ \bar{\e}_{2C} \g_{\m_7 ]} \e_{1C})\right]  \nonumber \\
  &+ &  12 d \ \left[ A^{(\a}_{[\m_2 \m_3} ( V^{\b)}_+ \ \bar{\e}_2 \g_{\m_4 \dots \m_7 ]} \e_1
  -  V^{\b)}_- \ \bar{\e}_{2C} \g_{\m_4 \dots \m_7 ]} \e_{1C})\right] \quad, \label{lambda7doubleprime}
  \end{eqnarray}
and the algebra closes provided that the coefficients are fixed to be
  \begin{eqnarray}
  & & c = \frac{21}{4} \quad , \qquad d = - \frac{7}{4} \quad ,
  \qquad e = -\frac{105}{8}  \quad , \nonumber \\
  & & f = -35 \quad , \qquad g = 70 \quad , \qquad \a = \frac{9}{4} \quad .
  \end{eqnarray}

In conclusion, the supersymmetry transformation for the eight-forms is
  \begin{eqnarray} \d
  A^{\a\b}_{\m_1 \dots \m_8} &=&
  -i \ V^\a_+ V^\b_+ \ \bar{\e} \g_{\m_1 \dots \m_8} \l_C
  + i \ V^\a_- V^\b_- \ \bar{\e}_C \g_{\m_1 \dots \m_8} \l \nonumber\\
  &+& 8 \ V^{(\a}_+ V^{\b )}_- \ \bar{\e} \g_{[\m_1 \dots \m_7} \psi_{\m_8 ]} -
  8 \ V^{(\a}_+ V^{\b )}_- \ \bar{\e}_C \g_{[\m_1 \dots \m_7} \psi_{\m_8 ]C} \nonumber \\
  &+& \tfrac{21}{4} A^{(\a}_{[\m_1 \dots \m_6} \d A^{\b)}_{\m_7 \m_8]}
  -\tfrac{7}{4} A^{(\a}_{[\m_1 \m_2} \d A^{\b)}_{\m_3 \dots \m_8]}
  - \tfrac{105i}{8} A^{(\a}_{[\m_1 \m_2} A^{\b)}_{\m_3 \m_4} \e_{\g\d} A^\g_{\m_5 \m_6} \d
  A^\d_{\m_7\m_8]}\nonumber \\
  & -& 35 A^{(\a}_{[\m_1 \m_2} A^{\b)}_{\m_3 \m_4} \d A_{\m_5 \dots \m_8 ]}
  +   70 A_{[\m_1 \dots \m_4} A^{(\a}_{\m_5 \m_6} \d A^{\b)}_{\m_7 \m_8]}
  \quad , \label{susy8final}
  \end{eqnarray}
while the gauge invariance of the field strengths
  \begin{equation}
  F^{\a\b}_{\m_1 \dots \m_9} = 9 \de_{[\m_1}
  A^{\a\b}_{\m_2 \dots \m_9]} + \tfrac{9}{4} F^{(\a}_{[\m_1 \dots \m_7} A^{\b)}_{\m_8\m_9]}
  - \tfrac{63}{4} F^{(\a}_{[\m_1 \dots \m_3 } A^{\b)}_{\m_4 \dots \m_9 ]}
  \end{equation}
requires
  \begin{equation}
  \d A^{\a\b}_{\m_1 \dots \m_8} = 8 \de_{[\m_1} \L^{\a\b}_{\m_2 \dots \m_8]} +\tfrac{1}{2}
  F^{(\a}_{[\m_1 \dots \m_7} \L^{\b)}_{\m_8]} - \tfrac{21}{2} F^{(\a}_{[\m_1 \dots \m_3}
  \L^{\b)}_{\m_4 \dots \m_8]} \quad .
  \end{equation}
Finally, the seven-form gauge parameter that appears in the commutator
is
  \begin{eqnarray}
  \L^{\a\b}_{\m_1 \dots \m_7} & =&
  A^{\a\b}_{\m_1 \dots \m_7 \s} \xi^\s
-\tfrac{1}{2} S^{\a\b} [ \bar{\e}_2
  \g_{\m_1 \dots \m_7} \e_1 - \bar{\e}_{2C}\g_{\m_1 \dots \m_7}
  \e_{1C} ] \nonumber \\
  &-&  \tfrac{21i}{8}
  A^\a_{[\m_1 \dots \m_6} (V^\b_+ \ \bar{\e}_2 \g_{\m_7 ]} \e_1
  +  V^\b_- \ \bar{\e}_{2C} \g_{\m_7 ]} \e_{1C} ) \nonumber \\
  &-& \tfrac{21}{8} A^\a_{[\m_1 \m_2} (
  V^\b_+ \ \bar{\e}_2 \g_{\m_3 \dots \m_7 ]} \e_1
  -  V^\b_- \ \bar{\e}_{2C} \g_{\m_3 \dots \m_7 ]} \e_{1C}) \quad ,
  \end{eqnarray}
as one obtains from eqs. (\ref{lambda7}), (\ref{lambda7prime}) and
(\ref{lambda7doubleprime}).

\section{Ten-forms\label{tenforms}}

The construction of ten-forms differs in an essential way from
that of the
six- and eight-forms: they do not have a field strength and therefore
they cannot be dual to some other form within the IIB theory.
They do not have propagating degrees of freedom,
since the charge associated to them must vanish.
Therefore there is no a priori limit on the number of ten-forms one could
introduce. Also the $SU(1,1)$ representations cannot be
guessed from the duality relations with lower rank forms.
However, their supersymmetry transformations are well defined.
We therefore proceed as before, determining the independent ten-forms
from the requirement that the supersymmetry algebra must close.
We want to determine the most general supersymmetry transformations for the ten-forms,
compatible with $U(1)$ invariance, for a given $SU(1,1)$
representation. We first prove that both a doublet and a quadruplet
of ten-forms are allowed, and then we discuss the claim that these
are the only possible ten-forms that are compatible with all the
symmetries of IIB supergravity.

\subsection{The doublet of ten-forms\label{double}}

We want to determine the supersymmetry transformations of a doublet
of ten-forms $A^\a_{\m_1 \dots \m_{10}}$ satisfying the reality
condition
  \begin{equation}
  (A^1 )^*_{\m_1 \dots \m_{10}} = A^2_{\m_1 \dots \m_{10}} \quad .
  \end{equation}
As we have seen already in the previous sections, the supersymmetry
transformation of any form consists of terms containing spinors,
plus possibly terms containing lower-rank forms and their
supersymmetry transformations. In the case of the ten-form doublet,
$U(1)$ invariance requires that the most general fermionic part in
the supersymmetry transformation of the ten-form doublet is
  \begin{eqnarray}
  \d A^\a_{\m_1\ldots\m_{10}}  & =&a \
  V^\a_- \ \bar{\e} \g_{\m_1 ...\m_{10}} \l + a^* \ V^\a_+ \ \bar{\e}_C
  \g_{\m_1 ...\m_{10}} \l_C
  \nonumber \\
  &+&  b \ V^\a_- \ \bar{\e}_C \g_{[\m_1 ... \m_9 }
  \psi_{\m_{10}]} - b^* \ V^\a_+ \ \bar{\e} \g_{[\m_1 ... \m_9 } \psi_{\m_{10}] C}
  \quad . \label{10formvar}
  \end{eqnarray}
The commutator of two such transformations contains the ten-form
gauge transformation
  \begin{eqnarray}
  \d A^\a_{\m_1 \dots \m_{10}} &=& 10 \de_{[\m_1} \L^\a_{\m_2\dots
  \m_{10}]} \nonumber\\
  &=& - 20 i \de_{[\m_1} \big( a \ V^\a_+ \ \bar{\e}_2 \g_{\m_2 \dots
  \g_{10}]} \e_{1C} +a^* \ V^\a_- \ \bar{\e}_{2C} \g_{\m_2 \dots
  \g_{10}]} \e_{1} \big) \quad ,
  \end{eqnarray}
provided that the coefficients $a$ and $b$ satisfy
  \begin{equation}
  b = 20 i a^* \quad .
  \end{equation}
Moreover, the additional terms in the commutator, containing the
five-form $F_5$ and the complex three-form $G_3$, vanish if $a$ is chosen
to be real.

In order to close the algebra, one also has to produce a general
coordinate transformation with parameter $\xi^\m$ (\ref{parameters}),
but this exactly cancels with the gauge transformation of
parameter\footnote{For lower rank $p$-forms these transformations
are obtained in the form $\xi^\rho F_{\rho\m_1\ldots \m_p}$, for
$p=D$ the vanishing of the $D+1$-form $F$ corresponds to the cancellation
of the two transformations. This result will be used again in the next subsection,
when we will
consider ten-forms in other representations of $SU(1,1)$.}
  \begin{equation}
  \L^{'\a}_{\m_1 \dots \m_9} = A^\a_{\m_1 \dots \m_9 \s} \xi^\s
  \quad .
  \end{equation}
As a result, the algebra closes without adding any term containing
lower-rank forms in the supersymmetry transformation of eq.
(\ref{10formvar}). Correspondingly, this ten-form doublet is
invariant with respect to the gauge transformations of the
lower-rank forms. Without loss of generality, we can fix
  \begin{equation}
  a=1 \quad ,
  \end{equation}
so that the resulting supersymmetry transformation for the ten-form
doublet is
  \begin{eqnarray}
  \d A^\a_{\m_1\ldots\m_{10}}  & =&
  V^\a_- \ \bar{\e} \g_{\m_1 ...\m_{10}} \l + V^\a_+ \ \bar{\e}_C
  \g_{\m_1 ...\m_{10}} \l_C
  \nonumber \\
  &+&  20 i \ V^\a_- \ \bar{\e}_C \g_{[\m_1 ... \m_9 }
  \psi_{\m_{10}]} + 20 i \ V^\a_+ \ \bar{\e} \g_{[\m_1 ... \m_9 } \psi_{\m_{10}] C}
  \quad . \label{10formvarfinal}
  \end{eqnarray}

\subsection{The quadruplet of ten-forms\label{quartet}}

We consider now a quadruplet of ten-forms $A^{\a\b\g}_{\m_1 \dots
\m_{10}}$, completely symmetric in $\a$, $\b$ and $\g$, and
satisfying the reality condition
  \begin{equation}
  (A^{111})^*_{\m_1 \dots \m_{10}} = A^{222}_{\m_1 \dots \m_{10}}
  \quad , \qquad (A^{112})^*_{\m_1 \dots \m_{10}} =
  A^{122}_{\m_1 \dots \m_{10}} \quad .
  \end{equation}
The most general supersymmetry transformation, compatible with the
reality condition and with $U(1)$ invariance, and consisting of
terms that only involve the spinors and terms containing the lower
rank forms and their supersymmetry transformations, is
  \begin{eqnarray}
  \d A_{\mu_1 \ldots \mu_{10}}^{\a\b\g} &=& a \  V^{(\a}_+ V^\b_+ V^{\g)}_- \
  \bar{\e}_C \g_{\mu_1 \ldots \mu_{10}} \l_C
  + a^* \  V^{(\a}_- V^\b_- V^{\g)}_+ \ \bar{\e} \g_{\mu_1 \ldots \mu_{10}} \l \nonumber\\
  &+& b \ V^{(\a}_+ V^\b_+ V^{\g)}_- \ \bar{\e} \g_{[ \mu_1 \ldots \mu_9} \psi_{\mu_{10}]C}
  - b^* \ V^{(\a}_- V^\b_- V^{\g)}_+ \ \bar{\e}_C \g_{[\mu_1 \ldots \mu_9} \psi_{\m_{10}]} \nonumber \\
  &+& c \ A_{[\mu_1 \ldots \mu_8}^{(\a\b} \d A_{\mu_9 \mu_{10}]}^{\g)} + d \ A_{[\mu_1 \mu_2}^{(\a}
  \d A_{\mu_3 \ldots \mu_{10}]}^{\b\g)} +e \ A^{(\a}_{[\mu_1 \ldots \mu_6}
  A^\b_{\mu_7 \mu_8} \d A^{\g)}_{\mu_9 \mu_{10}]} \label{10formquadrupletsusy}\\
  &+& f \ A^{(\a}_{[\mu_1 \mu_2} A^\b_{\mu_3 \mu_4} \d A^{\g)}_{\mu_5 \ldots \mu_{10}]}
  +  g \ A_{[\mu_1 \ldots \mu_4} A^{(\a}_{\mu_5 \mu_6} A^\b_{\mu_7 \mu_8} \d
  A^{\g)}_{\mu_9 \mu_{10}]} \nonumber \\
  &+& h \ A^{(\a}_{[\mu_1 \mu_2} A^\b_{\mu_3 \mu_4} A^{\g)}_{\mu_5 \mu_6} \d A_{\mu_7 \ldots \mu_{10}]}
  + i k \ A^{(\a}_{[\mu_1 \mu_2} A^\b_{\mu_3 \mu_4} A^{\g)}_{\mu_5 \mu_6}
  \e_{\d\t} A^\d_{\mu_7\mu_8} \d A^\t_{\mu_9 \mu_{10}]} \ .
  \nonumber
  \end{eqnarray}

We want to analyze the commutator of two such transformations.

We first consider the contribution coming from the fermionic terms,
{\ie}, the first two lines of eq. (\ref{10formquadrupletsusy}). Those
produce the ten-form gauge transformation
  \begin{eqnarray}
  & & \d A^{\a\b\g}_{\m_1 \dots \m_{10}} = 10 \de_{[\m_1}
  \L^{\a\b\g}_{ \m_2 \dots \m_{10}]} \nonumber \\
  && =\tfrac{20}{3}i \de_{[\m_1} ( a \ V^{(\a}_+ V^\b_+ V^{\g)}_- \
  \bar{\e}_2 \g_{\m_2 \dots \m_{10}]} \e_{1C} + a^* \ V^{(\a}_- V^\b_- V^{\g)}_+ \
  \bar{\e}_{2C} \g_{\m_2 \dots \m_{10}]} \e_{1} ) \label{lambda9}
  \end{eqnarray}
together with the terms
  \begin{eqnarray}
  & & \tfrac{20}{3}a  \ F^{(\a\b}_{[\m_1 \dots \m_9} \left( V^{\g)}_+ \ \bar{\e}_2
  \g_{\m_{10}]} \e_{1C} + V^{\g)}_- \ \bar{\e}_{2C}
  \g_{\m_{10}]} \e_{1} \right) \nonumber \\
  & & -20 a i \ S^{(\a\b} \left( \bar{\e}_2
  \g_{[\m_1 \dots \m_7} \e_1 - \bar{\e}_{2C}\g_{[\m_1 \dots \m_7}
  \e_{1C} \right) F^{\g)}_{\m_8 \m_9 \m_{10}]} \quad ,
  \label{10formextracomm}
  \end{eqnarray}
provided that
  \begin{equation}
  - \tfrac{20i}{3}  a = b
  \end{equation}
and $a$ is chosen to be imaginary. Without loss of generality, we
can fix
  \begin{equation}
  a = i
  \end{equation}
from now on. As in the case of the ten-form doublet of the previous
subsection, a general coordinate transformation is automatically
produced by means of a compensating gauge transformation of
parameter
  \begin{equation}
  \L^{'\a\b\g}_{\m_1 \dots \m_9} = A^{\a\b\g}_{\m_1 \dots \m_9 \s} \xi^\s
  \quad .\label{lambdaprime9}
  \end{equation}

We
assume that the ten-form quadruplet transforms non-trivially with
respect to the lower-rank form gauge transformations, and in
particular we make the Ansatz
  \begin{equation} \d A^{\a\b\g}_{\mu_1 \ldots \mu_{10}} = \a
  F^{(\a\b}_{[\mu_1 \ldots \mu_9} \L^{\g)}_{\mu_{10}]} + \b F^{(\a}_{[\mu_1 \mu_2
  \mu_3} \L_{\mu_4 \ldots \mu_{10}]}^{\b\g)} \quad .\label{10formgaugetransf}
  \end{equation}
We will comment on this choice at the end of this subsection. We now
proceed exactly as in the previous cases, considering the terms in
the commutator coming from the last  three lines of eq.
(\ref{10formquadrupletsusy}), as well as the two terms in eq.
(\ref{10formextracomm}). Those have to generate the gauge
transformations of eq. (\ref{10formgaugetransf}), possibly together
with an additional ten-form gauge transformation. The final result is
that the ten-form gauge transformation of parameter
  \begin{eqnarray}
  \L^{''\a\b\g}_{\m_1 \dots \m_9 } &=& - \tfrac{2i}{5} c \
  A^{(\a\b}_{[\m_1 \dots \m_8} \left( V^{\g)}_+ \ \bar{\e}_2
  \g_{\m_{9}]} \e_{1C} + V^{\g)}_- \ \bar{\e}_{2C}
  \g_{\m_{9}]} \e_{1} \right) \nonumber \\
  &-&\tfrac{2}{5} d \ A^{(\a}_{[\m_1 \m_2 } S^{\b\g)} \left( \bar{\e}_2
  \g_{\m_3 \dots \m_9 ]} \e_1 - \bar{\e}_{2C}\g_{\m_3 \dots \m_9 ]}
  \e_{1C} \right) \quad \label{lambdadoubleprime9}
  \end{eqnarray}
is produced, while  the coefficients are determined to be
  \begin{eqnarray}
  && \a = -\frac{2}{3} \quad , \qquad \b =32 \quad , \qquad c =-12\quad ,
  \nonumber \\
  & & d =3 \quad , \qquad e =-\frac{63}{4} \quad , \qquad f =
  \frac{21}{4} \quad , \nonumber \\
  & & g = -210 \quad , \qquad h = 105 \quad , \qquad k =
  \frac{315}{8} \quad . \label{coefficients10form}
  \end{eqnarray}

Summarizing, the supersymmetry transformation of the ten-form
quadruplet is
  \begin{eqnarray}
  \d A_{\mu_1 \ldots \mu_{10}}^{\a\b\g} &=& i \  V^{(\a}_+ V^\b_+ V^{\g)}_- \
  \bar{\e}_C \g_{\mu_1 \ldots \mu_{10}} \l_C
  -i \  V^{(\a}_- V^\b_- V^{\g)}_+ \ \bar{\e} \g_{\mu_1 \ldots \mu_{10}} \l
\nonumber\\
  &+& \tfrac{20}{3} \ V^{(\a}_+ V^\b_+ V^{\g)}_- \ \bar{\e} \g_{[ \mu_1 \ldots \mu_9} \psi_{\mu_{10}]C}
  - \tfrac{20}{3} \ V^{(\a}_- V^\b_- V^{\g)}_+ \ \bar{\e}_C \g_{[\mu_1 \ldots \mu_9} \psi_{\m_{10}]} \nonumber \\
  &-& 12 \ A_{[\mu_1 \ldots \mu_8}^{(\a\b} \d A_{\mu_9 \mu_{10}]}^{\g)} + 3 \ A_{[\mu_1 \mu_2}^{(\a}
  \d A_{\mu_3 \ldots \mu_{10}]}^{\b\g)} -\tfrac{63}{4} \ A^{(\a}_{[\mu_1 \ldots \mu_6}
  A^\b_{\mu_7 \mu_8} \d A^{\g)}_{\mu_9 \mu_{10}]} \label{10formquadrupletsusyfinal}\\
  &+& \tfrac{21}{4} \ A^{(\a}_{[\mu_1 \mu_2} A^\b_{\mu_3 \mu_4} \d A^{\g)}_{\mu_5 \ldots \mu_{10}]}
  -  210 \ A_{[\mu_1 \ldots \mu_4} A^{(\a}_{\mu_5 \mu_6} A^\b_{\mu_7 \mu_8} \d
  A^{\g)}_{\mu_9 \mu_{10}]} \nonumber \\
  &+& 105 \ A^{(\a}_{[\mu_1 \mu_2} A^\b_{\mu_3 \mu_4} A^{\g)}_{\mu_5 \mu_6} \d A_{\mu_7 \ldots \mu_{10}]}
  +  \tfrac{315i}{8} \ A^{(\a}_{[\mu_1 \mu_2} A^\b_{\mu_3 \mu_4} A^{\g)}_{\mu_5 \mu_6}
  \e_{\d\t} A^\d_{\mu_7\mu_8} \d A^\t_{\mu_9 \mu_{10}]} \ ,
  \nonumber
  \end{eqnarray}
while its gauge transformation is
  \begin{equation}
  \d A^{\a\b\g}_{\m_1 \dots \m_{10}} = 10 \de_{[\m_1}
  \L^{\a\b\g}_{ \m_2 \dots \m_{10}]} -\tfrac{2}{3}
  F^{(\a\b}_{[\mu_1 \ldots \mu_9} \L^{\g)}_{\mu_{10}]} + 32 F^{(\a}_{[\mu_1 \mu_2
  \mu_3} \L_{\mu_4 \ldots \mu_{10}]}^{\b\g)} \quad .
  \label{10formquadrupletgaugefinal}
  \end{equation}
Finally, the ten-form gauge transformation parameter appearing in the
supersymmetry algebra is
  \begin{eqnarray}
  \L^{\a\b\g}_{\m_1 \dots \m_9} &=&
  A^{\a\b\g}_{\m_1 \dots \m_9 \s} \xi^\s
  -\tfrac{2}{3} (V^{(\a}_+ V^\b_+ V^{\g)}_- \
  \bar{\e}_2 \g_{\m_1 \dots \m_{9}} \e_{1C} -  V^{(\a}_- V^\b_- V^{\g)}_+ \
  \bar{\e}_{2C} \g_{\m_1 \dots \m_{9}} \e_{1} )
  \nonumber \\
  &+& \tfrac{24i}{5}  \
  A^{(\a\b}_{[\m_1 \dots \m_8} \left( V^{\g)}_+ \ \bar{\e}_2
  \g_{\m_{9}]} \e_{1C} + V^{\g)}_- \ \bar{\e}_{2C}
  \g_{\m_{9}]} \e_{1} \right) \nonumber \\
  &-&\tfrac{6}{5} \ A^{(\a}_{[\m_1 \m_2 } S^{\b\g)} \left( \bar{\e}_2
  \g_{\m_3 \dots \m_9 ]} \e_1 - \bar{\e}_{2C}\g_{\m_3 \dots \m_9 ]}
  \e_{1C} \right) \quad ,
  \end{eqnarray}
as it results from eqs. (\ref{lambda9}), (\ref{lambdaprime9}) and
(\ref{lambdadoubleprime9}).

To conclude this subsection, we want to comment on the bosonic gauge
transformation of eq. (\ref{10formquadrupletgaugefinal}). Even though
the supersymmetry algebra restricts us in our case to ten dimensions,
it turns out that
the bosonic gauge algebra closes for {\it arbitrary} dimension.
In particular one can write down an eleven-form field strength
 that is gauge invariant with respect to
a gauge transformation of the form (\ref{10formgaugetransf}):
  \begin{equation}
  F^{\a\b\g}_{\m_1 \dots \m_{11}} = 11 \de_{[\m_1 } A^{\a\b\g}_{\m_2
  \dots \m_{11}]} + \tfrac{11}{2} \a \ A^{(\a}_{[\m_1 \m_2} F^{\b\g)}_{\m_3
  \dots \m_{11}]} + \tfrac{11}{8}\b \ A^{(\a\b}_{[\m_1 \dots \m_8}
  F^{\g)}_{\m_9 \m_{10} \m_{11}]} \quad ,\label{11formfieldstrength}
  \end{equation}
where the coefficients $\a$ and $\b$ have to satisfy the constraint
  \begin{equation}
  \b = -48 \a \quad .
  \end{equation}
This relation is in agreement with the values of $\a$ and $\b$
given in eq. (\ref{coefficients10form}) and obtained imposing
supersymmetry.
This suggests that the bosonic gauge algebra has
an underlying structure that is independent of supersymmetry in ten 
dimensions\footnote{This type of gauge algebra is also observed in the
doubled fields approach, see \cite{Juliaetal}.}.

\subsection{Other ten-forms?\label{more}}

We now want to show that no other ten-forms can be included in the
supersymmetry algebra of IIB supergravity. In order to do this,
we consider the most general Ansatz for the supersymmetry
transformation of a ten-form in a generic representation of
$SU(1,1)$. Without loss of generality, we can limit ourselves to
ten-forms with vanishing $U(1)$-charge.
The simplest such example is a singlet of
$SU(1,1)$, for which the supersymmetry transformation necessarily is
    \begin{equation}
    \d A_{\m_1 \dots \m_{10}} = \bar{\e} \g_{[\m_1 \dots \m_9 }
    \psi_{\m_{10}]} + \bar{\e}_C \g_{[\m_1 \dots \m_9 }
    \psi_{\m_{10}]C} \quad .
    \end{equation}
The commutator of two such transformations closes.
This is not surprising since $A_{(10)}$
is the volume form,
  \begin{equation}
  A_{\m_1 \dots \m_{10}} \propto \e_{\m_1 \ldots \m_{10}} =
  e_{\m_1}{}^{a_1} \ldots e_{\m_{10}}{}^{a_{10}} \e_{a_1 \ldots
  a_{10}} \quad .
  \end{equation}
This means that
there are no {\sl independent} ten-form singlets in the supersymmetry
algebra of IIB.

One could ask whether additional ten-form doublets could result
from objects of the form $A^{\a_1 \dots \a_{2n+1}}_{\m_1 \dots
\m_{10}}$, when $2n$ $SU(1,1)$ indices are pairwise antisymmetrized.
However, because of the constraint of eq. (\ref{VVe}) these forms
are the same as the one we obtained in section \ref{double}, and therefore there is only a
single doublet of ten-forms in the theory. This argument can be
iterated, so that for each object with an odd number of $SU(1,1)$
indices, only the components in the completely symmetric
representation are independent of the ten-forms
belonging to lower representations.

Therefore, given a ten-form with $n$ $SU(1,1)$ indices,
one has to consider only the completely symmetric $SU(1,1)$
representation. Let us consider the case $n=2$ first.
The most general Ansatz for the fermionic terms is
\be
   \d A^{\a\b}_{(10)} = a_1 V^{(\a}_+ V^{\b)}_- \bar{\e} \g_{(9)}
    \psi + a_2 V^{(\a}_+ V^{\b)}_- \bar{\e}_C \g_{(9)}
    \psi_{C} + b_1 V^{\a}_+ V^{\b}_+ \bar{\e} \g_{(10)}
    \l_{C} + b_2 V^{\a}_- V^{\b}_- \bar{\e}_C \g_{(10)} \l  \quad .
\ee As in the case of the singlet, one can close the algebra on this
Ansatz, but again it is not an independent field. It turns out to be
the variation of a composite field:
  \be
  \d \left( \tfrac{1}{2} S^{\a\b} \e_{(10)} \right) =
  \d \left( V^{(\a}_+ V^{\b)}_- \e_{(10)} \right) = V^{(\a}_+ V^{\b)}_- \d \e_{(10)}
  + \d V^{(\a}_+ V^{\b)}_- \e_{(10)} + V^{(\a}_+ \d V^{\b)}_- \e_{(10)} \,.
\ee This generalises to ten-forms with $n = 2m$ $SU(1,1)$-indices,
for which we can also close the algebra, but end up with the
variation of the composite field \be S^{(\a_1\b_1} \ldots S^{\a_{m}
\b_{m})} \e_{(10)} \quad . \ee The case of $n$ odd is different,
since the requirement of vanishing $U(1)$ charge does not allow one
to write down a volume form. In this case the Ansatz for the
fermionic part of the supersymmetry transformation is (we set here
$n=2m+1$)
  \begin{eqnarray}
  \d A_{\mu_1 \ldots \mu_{10}}^{\a_1 \dots \a_{2m+1}} &=&
  a \  V^{(\a_1}_+ \dots V^{\a_{m+1}}_+
  V^{\a_{m+2}}_- \dots V^{\a_{2m+1})}_- \
  \bar{\e}_C \g_{\mu_1 \ldots \mu_{10}} \l_C \nonumber \\
  &+& a^* \   V^{(\a_1}_- \dots V^{\a_{m+1}}_-
  V^{\a_{m+2}}_+ \dots V^{\a_{2m+1})}_+ \ \bar{\e} \g_{\mu_1 \ldots \mu_{10}} \l \nonumber\\
  &+& b \ V^{(\a_1}_+ \dots V^{\a_{m+1}}_+
  V^{\a_{m+2}}_- \dots V^{\a_{2m+1})}_- \ \bar{\e} \g_{[ \mu_1 \ldots \mu_9}
  \psi_{\mu_{10}]C} \nonumber \\
  &-& b^* \  V^{(\a_1}_- \dots V^{\a_{m+1}}_-
  V^{\a_{m+2}}_+ \dots V^{\a_{2m+1})}_+
\ \bar{\e}_C \g_{[\mu_1 \ldots \mu_9} \psi_{\m_{10}]}
  \quad . \label{10formtransf2m+1plet}
  \end{eqnarray}
It can be shown that only for the case $m=0$, {\ie}, the
doublet that we already considered, the commutator of two such
transformations closes producing just a ten-form gauge transformation
and a general coordinate transformation. As we have seen already for
the quadruplet ($m=1$), extra terms are
generated that need to combine with additional terms in eq.
(\ref{10formtransf2m+1plet}), containing lower-rank forms and their
supersymmetry transformations, to produce bosonic gauge
transformations. An explicit analysis shows that these terms can only
be written for the quadruplet. Higher $SU(1,1)$
representations require introducing additional contributions from the scalars in
these bosonic terms, and the supersymmetry commutator
produces derivatives of these scalars. These contributions can not be identified with any
parameter that appears in the supersymmetry algebra. This suggests that
only a doublet and a quadruplet can be consistently included in the
supersymmetry algebra of IIB.

\section{The complete IIB transformation rules and algebra} \label{results}

This section collects our results for the $SU(1,1)$-democratic version of $D=10$ IIB supergravity.
We present the supersymmetry transformation rules, the transformation rules of the
$p$-forms under bosonic gauge transformations, the definition of gauge invariant curvatures,
and finally the results for the commutator of two supersymmetry transformations. Of course all the transformations
and definitions are interdependent. All results have been derived only up to the
quadratic order in the fermions.

The supersymmetry transformation rules in Einstein frame, in the notation of
\cite{SW, schwarz}, are:
\begin{eqnarray}
  \d e_\m{}^a &=& i \bar{\e} \g^a \psi_\m
      + i \bar{\e}_C \g^a \psi_{\m C}  \quad ,
\label{res-e}
\\
  \d \psi_\m &=& D_\m \e +\tfrac{i}{480} F_{\m\n_1 ...\n_4 } \g^{\n_1 ...\n_4 } \e
  +\tfrac{1}{96} G^{\n\r\s} \g_{\m\n\r\s} \e_C
  -\tfrac{3}{32}
  G_{\m\n\r} \g^{\n\r} \e_C \quad ,
\label{res-psi}
\\
  \d A^\a_{\m\n} &=& V^\a_- \ \bar{\e} \g_{\m\n} \l
                      +V^\a_+ \ \bar{\e}_C \g_{\m\n} \l_C
  +4i V^\a_- \ \bar{\e}_C \g_{[\m} \psi_{\n ]}
  +4i V^\a_+ \ \bar{\e} \g_{[\m} \psi_{\n ]C}
  \quad ,
\label{res-A2}
\\
  \d A_{\m\n\r\s} &=&\bar{\e} \g_{[ \m\n\r} \psi_{\s ]}
                      -\bar{\e}_C \g_{[ \m\n\r} \psi_{\s ]C}
  -\tfrac{3i}{8} \e_{\a\b} A^\a_{[\m\n} \d A^\b_{\r \s ]} \quad ,
\label{res-A4}
\\
  \d \l &=& i P_\m \g^\m \e_C -\tfrac{i}{24}
  G_{\m\n\r}^{} \g^{\m\n\r} \e \quad ,
\label{res-l}
\\
  \d V^\a_+ &=& V^\a_- \ \bar{\e}_C \l  \quad ,
\label{res-Vm}
\\
  \d V^\a_- &=& V^\a_+ \ \bar{\e} \l_C   \quad ,
\label{res-Vp}
\\
\d A^\a_{\m_1 \ldots \m_6} &=&
  i V^\a_- \bar{\e} \g_{\m_1\ldots\m_6} \l
- i V^\a_+ \bar{\e}_C \g_{\m_1\ldots\m_6} \l_C
\nonumber\\
&& + 12 \left(V^\a_- \bar{\e}_C \g_{[\m_1\ldots\m_5 } \psi_{\m_6]}
          - V^\a_+ \bar{\e} \g_{[\m_1\ldots\m_5 } \psi_{C\,\m_6] }\right)
\nonumber \\
 &&+ 40 A_{[\m_1\ldots\m_4} \d A^\a_{\m_5\m_6]}
    -20 \d A_{[\m_1\ldots\m_4} A^\a_{\m_5\m_6]}
\nonumber\\
 && -\tfrac{15i}{2} A^\a_{[\m_1\m_2} \e_{\b\g} A^\b_{\m_3\m_4} \d A_{\m_5\m_6]}^\g\quad ,
\label{res-A6}
\\
\d A^{\a\b}_{\m_1 \ldots \m_8} &=&
  +i V^{(\a}_- V^{\b)}_- \bar{\e}_C \g_{\m_1\ldots\m_8} \l
 - i V^{(\a}_+ V^{\b)}_+ \bar{\e}   \g_{\m_1\ldots\m_8} \l_C
\nonumber\\
 &&+ 8 V^{(\a}_+V^{\b)}_- (\bar{\e} \g_{[\m_1\ldots\m_7} \psi_{\m_8]}
   -  \bar{\e}_C \g_{[\m_1\ldots\m_7} \psi_{C\,\m_8]})
\nonumber \\
 &&+\tfrac{21}{4} A^{(\a}_{[\m_1\ldots \m_6} \d A^{\b)}_{\m_7\m_8]}
  - \tfrac{7}{4}  A^{(\a}_{[\m_1\m_2}        \d A^{\b)}_{\m_3\ldots\m_8]}
\nonumber \\
 &&
  -35 A^{(\a}_{[\m_1\m_2} A^{\b)}_{\m_3\m_4} \d A_{\m_5\ldots\m_8]}
 + 70 A_{[\m_1\ldots\m_4} A^{(\a}_{\m_5\m_6} \d A^{\b)}_{\m_7\m_8]}
\nonumber\\
&& -  \tfrac{105i}{8} A^{(\a}_{[\m_1\m_2} A^{\b)}_{\m_3\m_4}
   \e_{\g\d} A^\g_{\m_5\m_6} \d A^\d_{\m_7\m_8]} \quad ,
\label{res-A8}
\\
\d A^{\a}_{\m_1 \ldots \m_{10}} &=&
   V^\a_- \bar{\e}  \g_{\m_1\ldots\m_{10}}\l
                       + V^\a_+ \bar{\e}_C\g_{\m_1\ldots\m_{10}}\l_C
\nonumber \\
 &&+  20i \left( V^\a_+ \bar{\e} \g_{[\m_1\ldots \m_9} \psi_{C\,\m_{10}]}
  + V^\a_- \bar{\e}_C \g_{[\m_1\ldots\m_9} \psi_{\m_{10}]}\right) \quad ,
\label{res-A10-2}
\\
\d A^{\a\b\g}_{\m_1 \ldots \m_{10}} &=&
  i V^{(\a}_+ V^\b_+ V^{\g)}_-\bar{\e}_C \g_{\mu_1 \ldots \mu_{10}} \l_C
- i V^{(\a}_- V^\b_- V^{\g)}_+ \bar{\e}  \g_{\mu_1 \ldots \mu_{10}} \l
\nonumber\\
  &&
  + \tfrac{20}{3}(
    V^{(\a}_+ V^\b_+ V^{\g)}_- \bar{\e}  \g_{[\mu_1 \ldots \mu_9} \psi_{C\,\mu_{10}]}
  - V^{(\a}_- V^\b_- V^{\g)}_+ \bar{\e}_C\g_{[\mu_1 \ldots \mu_9} \psi_{\m_{10}]}
    )
\nonumber \\
&& - 12 \,  A_{[\mu_1 \ldots \mu_8}^{(\a\b} \d A_{\mu_9 \mu_{10}]}^{\g)}
   + 3  \, A_{[\mu_1 \mu_2}^{(\a}  \d A_{\mu_3 \ldots \mu_{10}]}^{\b\g)}
\nonumber \\
&& - \, \tfrac{63}{4} A^{(\a}_{[\mu_1 \ldots \mu_6} A^\b_{\mu_7 \mu_8}
                     \d A^{\g)}_{\mu_9 \mu_{10}]}
   + \tfrac{21}{4} \, A^{(\a}_{[\m_1 \m_2} A^\b_{\mu_3 \mu_4}
                     \d A^{\g)}_{\mu_5 \ldots \mu_{10}]}
\nonumber \\
&& - \, 210 A_{[\mu_1 \ldots \mu_4} A^{(\a}_{\mu_5 \mu_6}
                      A^\b_{\m_7 \m_8} \d A^{\g)}_{\m_9 \m_{10}]}
 +105 A^{(\a}_{[\m_1 \m_2} A^\b_{\m_3 \m_4} A^{\g)}_{\m_5 \m_6}
   \d A_{\m_7 \ldots \m_{10}]}
\nonumber \\
&& +\tfrac{315i}{8} \, A^{(\a}_{[\mu_1 \mu_2} A^\b_{\mu_3 \mu_4}
    A^{\g)}_{\mu_5 \mu_6} \e_{\d\t} A^\d_{\mu_7\mu_8} \d A^\t_{\mu_9 \mu_{10}]}
\quad .
\label{res-A10-4}
\end{eqnarray}
For the bosonic gauge-transformations we find:
\begin{eqnarray}
\d A^\a_{\m_1\m_2} &=& 2 \de_{[\m_1} \L^\a_{\m_2]}\quad ,
\\
\d A_{\m_1\ldots\m_4} &=& 4 \de_{[\m_1} \L_{\m_2\m_3\m_4]}
  - \tfrac{i}{4} \e_{\g\d} \L^\g_{[\m_1}F^\d_{\m_2\m_3\m_4]}\quad ,
\\
\d A^\a_{\m_1\ldots\m_6} &=& 6 \de_{[\m_1} \L^\a_{\m_2\ldots\m_6]}
     - 8 \L^\a_{[\m_1} F_{\m_2\ldots\m_6]}
  - \tfrac{160}{3} F^\a_{[\m_1\m_2\m_3}\L_{\m_4\m_5\m_6]}\quad ,
\\
\d A^{\a\b}_{\m_1\ldots\m_8} &=& 8 \de_{[\m_1} \L^{(\a\b)}_{\m_2\ldots\m_8]}
     +\tfrac{1}{2} F^{(\a}_{[\m_1\ldots\m_7} \L^{\b)}_{\m_8]}
  - \tfrac{21}{2} F^{(\a}_{[\m_1\m_2\m_3} \L^{\b)}_{\m_4\ldots\m_8]}\quad ,
\\
\d A^{\a}_{\m_1 \ldots \m_{10}} &=& 10 \de_{[\m_1} \L^{\a}_{\m_2\ldots\m_{10}]}\quad ,
\\
\d A^{\a\b\g}_{\m_1 \ldots \m_{10}} &=&
   10 \de_{[\m_1} \L^{(\a\b\g)}_{\m_2\ldots\m_{10}]}
   -\tfrac{2}{3} F^{(\a\b}_{[\m_1 \ldots \m_9} \L^{\g)}_{\m_{10}]}
  + 32 F^{(\a}_{[\m_1 \m_2 \m_3} \L_{\m_4 \ldots \m_{10}]}^{\b\g)} \quad .
\end{eqnarray}
For the $p$-form fields we define field-strengths invariant under the bosonic gauge
transformations
\begin{eqnarray}
F^\a_{\mu_1\mu_2\mu_3} &=& 3 \de_{[\mu_1} A^\a_{\mu_2 \mu_3]} \quad ,
\\
F_{\mu_1 \ldots \mu_5} &=& 5 \de_{[\m_1} A_{\m_2\ldots\m_5 ]}
   + \tfrac{5i}{8} \e_{\a\b} A^\a_{[\m_1\m_2} F^{\b}_{\m_3\m_4\m_5 ]} \quad ,
\\
F^{\a}_{\mu_1 \ldots \mu_7} &=& 7 \de_{[\m_1} A^\a_{\m_2 \ldots \m_7]}
     + 28 A^\a_{[\m_1\m_2} F_{\m_3\ldots\m_7]}
  - \tfrac{280}{3} F^\a_{[\m_1\m_2\m_3}A_{\m_4\ldots\m_7]} \quad ,
\\
F^{\a\b}_{\mu_1 \ldots \mu_9} &=& 9 \de_{[\m_1} A^{\a\b}_{\m_2\ldots\m_9]} +
  \tfrac{9}{4} F^{(\a}_{[\m_1\ldots\m_7} A^{\b)}_{\m_8\m_9]}
 - \tfrac{63}{4} F^{(\a}_{[\m_1\m_2\m_3}A^{\b)}_{\m_4\ldots\m_9]} \quad ,
\\
F^{\a}_{\mu_1 \ldots \mu_{11}} &=&
 11 \partial_{[\mu_1} A^{\a}_{\mu_2 \ldots \mu_{11}]} = 0\quad ,
\\
F^{\a\b\g}_{\mu_1 \ldots \mu_{11}} &=&
  11 ( \partial_{[\mu_1} A^{\a\b\g}_{\mu_2 \ldots \mu_{11}]}
 -\tfrac{1}{3} F^{(\a \b}_{[\mu_1 \ldots \mu_9} A^{\g)}_{\mu_{10} \mu_{11}]}
+ 4  F^{(\a}_{[\mu_1 \mu_2 \mu_3} A^{\b \g)}_{\mu_4 \ldots \mu_{11}[}) = 0 \quad .
\end{eqnarray}
The duality relations between these field-strengths are:
\begin{eqnarray}
F^\a_{\m_1\ldots\m_7} &=&
  -\tfrac{i}{3!} \e_{\m_1\ldots\m_7 \m\n\r} S^{\a\b} \e_{\b\g} F^{\g ; \m\n\r}\quad ,
\\
F^{\a\b}_{\m_1\ldots\m_9} &=&
   i \e_{\m_1\ldots\m_9}{}^\r [ V^\a_+ V^\b_+ P^*_\r - V^\a_- V^\b_- P_\r ] \quad .
\end{eqnarray}

The commutator of two supersymmetry transformations, $[\d(\e_1),\d(\e_2)]$ must close on
symmetry transformations of the IIB multiplet. In fact, as we saw in previous sections,
this is the way the results of this paper have been obtained. We find the following
contributions to $[\d(\e_1),\d(\e_2)]$:
\begin{eqnarray}
\xi^\m &=&
   i \bar{\e}_2 \g^\m \e_1 +i \bar{\e}_{2C} \g^\m \e_{1C}\quad ,
\\
\L^\a_\m &=&
   A^\a_{\m \n} \xi^\n
 -2i [ V_+^\a \bar{\e}_2 \g_\m \e_{1C}
         + V_-^\a \bar{\e}_{2C} \g_\m \e_1 ] \quad ,
\\
\L_{\m_1\m_2\m_3} &=&
   A_{\m_1 \m_2 \m_3 \n} \xi^\n +
  \tfrac{1}{4}[ \bar{\e}_2 \g_{\m_1\m_2\m_3} \e_1
    - \bar{\e}_{2C} \g_{\m_1\m_2\m_3} \e_{1C} ] \quad ,
\\
\L^\a_{\m_1\ldots\m_5} &=&
    A^\a_{\m_1\ldots\m_5 \r} \xi^\r
 -2 V^\a_- \bar{\e}_{2C} \g_{\m_1\dots\m_5} \e_1
    +2 V^\a_+ \bar{\e}_2 \g_{\m_1\ldots\m_5}\e_{1C}
\nonumber\\
&&    +
    \tfrac{40}{3} A_{[\m_1\ldots\m_4} \L^\a_{\m_5]}
  -\tfrac{40}{3} \L_{[\m_1\m_2\m_3} A^\a_{\m_4\m_5]} \quad ,
\\
\L^{\a\b}_{\m_1\ldots\m_7} &=&
   A^{(\a\b)}_{\m_1 \ldots \m_7 \nu} \xi^\nu
 - 2 V^{(\a}_+V^{\b)}_- \left(\bar{\e}_{2} \g_{\m_1\ldots\m_7}\e_{1}
   -  \bar{\e}_{2C} \g_{\m_1\ldots\m_7}\e_{1C}\right)
\nonumber\\
&& + \tfrac{21}{16} A_{[\m_1\ldots \m_6}^{(\a}\L_{\m_7]}^{\b)}
  - \tfrac{21}{16}A_{[\m_1\m_2}^{(\a}\L_{\m_3 \ldots \m_7]}^{\b)}\quad ,
\\
\L^{\a}_{\m_1\ldots\m_9} &=& -2 i \left( V^\a_+
  \bar{\e}_2 \g_{\m_1 \dots \m_9} \e_{1C} + \ V^\a_-
  \bar{\e}_{2C} \g_{\m_1 \dots \m_9} \e_{1} \right)
\\
\L^{\a\b\g}_{\m_1\ldots\m_9} &=&
 A^{(\a\b\g)}_{\m_1\ldots\m_9\n}\xi^\n
    -\tfrac{2}{3}\left( V^{(\a}_+ V^\b_+ V^{\g)}_- \bar{\e}_2 \g_{\m_1\ldots\m_9}\e_{1C}
                 - V^{(\a}_-V^\b_-V^{\g)}_+   \bar{\e}_{2C} \g_{\m_1\ldots\m_9}\e_{1}\right)
\nonumber\\
&& + \tfrac{24i}{5}A^{(\a\b}_{[\m_1 \ldots\m_8}
   (V_+^{\g)} \bar{\e}_2 \g_{\m_9]} \e_{1C}
         + V_-^{\g)} \bar{\e}_{2C} \g_{\m_9]} \e_1)
\nonumber\\
&&  - \tfrac{6}{5}A^{(\a}_{[\m_1\m_2} S^{\b\g)}
(  \bar{\e}_2 \g_{\m_3\ldots\m_9]} \e_1
    - \bar{\e}_{2C} \g_{\m_3\ldots\m_9]} \e_{1C})
\quad .
\end{eqnarray}
This concludes the summary of our main results. In the next section we will
present the IIB supergravity multiplet in a real formulation in both
Einstein frame and string frame.

\section{$U(1)$ gauge fixing and string frame\label{SF}}

The results we derived so far were in Einstein frame. To go to string frame we
will first choose a $U(1)$ gauge,
so that the dependence on the dilaton becomes explicit.
Our choice is
\begin{equation}
V_-^1 \in \mathbb{R} \Rightarrow V_+^2 = V_-^1 \quad .
\label{U1-gauge}
\end{equation}
To preserve this condition the supersymmetry transformations have to be modified by a
field dependent $U(1)$ gauge transformation:
\begin{equation}
   \delta'(\epsilon)=\delta(\epsilon)
  + \delta_{U(1)}\left(\frac{i}{2V_-^1}(V_-^2\bar\epsilon_C\lambda -
                              V_+^1\bar\epsilon\lambda_C)\right)\quad .
\end{equation}
This modification is only visible on the scalars since
on the fermions it gives rise to terms cubic in fermionic variables.
The $SU(1,1)$ transformations are also modified:
the condition (\ref{U1-gauge}) is preserved under a
combination of an $SU(1,1)$ and a $U(1)$ transformation. On a field $\chi$
of $U(1)$-charge $q$ the required $U(1)$ transformation is
\begin{equation}
   \chi \to e^{iq\theta}\chi\,,\
   {\rm with}\
   e^{2i\theta} = \frac{\a+\b z}{\bar\a +\bar\b \bar z}\quad ,
\label{compU1}
\end{equation}
where the coordinate $z$ is
defined in (\ref{defz}).
This is of course visible on all fermions.

To make the dilaton and axion explicit we set
\begin{equation}
   V_-^1=V_+^2=\frac{1}{\sqrt{1-z\bar{z}}}\,,\quad
   V_-^2=(V_+^1)^*=\frac{z}{\sqrt{1-z\bar{z}}}\quad .
\end{equation}
Using (\ref{deftau}) and (\ref{deftlf}) we find (we now drop the prime on the redefined
supersymmetry transformation)
\begin{equation}
   \delta\tau = -2ie^{-\phi}e^{-2i\Lambda}
   \bar\epsilon\lambda_C\quad ,
\end{equation}
where
\begin{equation}
   e^{-2i\Lambda} = \frac{1-i\tau}{1+i\bar{\tau}}\quad .
\end{equation}
Useful variables are
\begin{equation}
   P_\mu = -\frac{i}{2}e^{-\phi}e^{-2i\Lambda}\partial_\mu\tau\,,
\qquad
   Q_\mu = \frac{1}{4}e^\phi\left(
   \frac{1-i\bar{\tau}}{1-i\tau}\partial_\m\tau +
   \frac{1+i\tau}{1+i\bar{\tau}}\partial_\m\bar\tau \right)\quad .
\end{equation}
It is convenient to get rid of the factors of $e^{-2i\Lambda}$ in
the supersymmetry transformation rules \cite{BHO}. To do this we
redefine the fermions by phase factors according to their $U(1)$
weights:
\begin{equation}
   \lambda' = e^{3i\Lambda/2}\lambda,\qquad
   \psi'_\mu = e^{i\Lambda/2}\psi_\mu,\qquad
   \e' = e^{i\Lambda/2} \e\quad .
\label{redeffermions}
\end{equation}
In the transformation rules the scalars $V_\pm^\alpha$ will now
occur everywhere in the combination $V_\pm^\alpha e^{\pm i \Lambda}$,
which are:
\begin{eqnarray}
   V_-^1 e^{-i\L} &=& \frac{1}{2}e^{\phi/2}(1-i\tau)\quad ,
\nonumber
\\
   V_+^1 e^{i\L} &=& \frac{1}{2}e^{\phi/2}(1-i\bar\tau)\quad ,
\nonumber
\\
   V_-^2 e^{-i\L} &=& \frac{1}{2}e^{\phi/2}(1+i\tau)\quad ,
\label{redefV}
\\
   V_+^2 e^{i\L} &=& \frac{1}{2}e^{\phi/2}(1+i\bar\tau)\quad .
\nonumber
\end{eqnarray}
Note that interchanging $V^1\leftrightarrow V^2$ corresponds to
$\tau\leftrightarrow -\tau$, $V_+\leftrightarrow V_-$ to
$\tau\leftrightarrow \bar\tau$. The transformation rules for the IIB
supergravity multiplet of \cite{SW, schwarz} now become \cite{bdgpt}:
\begin{eqnarray}
\d e_\m{}^a &=& i (\bar{\e} \g^a \psi_\m ) + {\rm h.c. } \label{Evielbein}\\
\d \psi_\m &=&  D_\m \e - \tfrac{i}{4}e^{\phi}\e\partial_\m\ell +
     \tfrac{i}{480} F_{\m_1\ldots\m_5 } \g^{\m_1\ldots\m_5}\gamma_\mu \e
\nonumber\\
 &&- \tfrac{1}{192}e^{\phi/2}
   \left( \g_{\m} \g^{\n\r\s} + 2 \g^{\n\r\s}\g_{\m}\right) \e_C
   (F^1-F^2+i\bar\tau(F^1+F^2))_{\n\r\s}\quad ,
\\
\delta A^1_{\m\n} &=&
  2i e^{\phi/2} \big\{
   (1-i\bar\tau)\,(\bar\e\g_{[\m}\psi_{C\,\n]}-\tfrac{i}{4}\bar\e_C\g_{\m\n}\l_C)
\nonumber\\
&&\phantom{2i e^{\phi/2} \big\{}
    +(1-i\tau)\,(\bar\e_C\g_{[\m}\psi_{\n]}-\tfrac{i}{4}\e\g_{\m\n]}\l) \big\} \quad ,
\label{dA1EF}
\\
\delta A^2_{\m\n} &=&
  2i e^{\phi/2} \big\{
   (1+i\bar\tau)\,(\bar\e\g_{[\m}\psi_{C\n]}-\tfrac{i}{4}\bar\e_C\g_{\m\n}\l_C)
\nonumber\\
&&\phantom{2i e^{\phi/2} \big\{}
      +(1+i\tau)\,(\bar\e_C\g_{[\m}\psi_{\n]}-\tfrac{i}{4}\e\g_{\m\n]}\l) \big\} \quad ,
\label{dA2EF}
\\
\d A_{\m\n\r\s}&=&
    \bar\e\g_{[\m\n\r}\psi_{\s]} +\ {\rm h.c.}
   -\tfrac{3i}{8}\e_{\a\b}A^\a_{[\m\n}\d A^\b_{\r\s]}\ ,
\\
\d \lambda &=& \tfrac{1}{2} e^\phi \g^\m \e_C \partial_\m\bar\tau
  -\tfrac{i}{48}e^{\phi/2}\g^{\n\r\s} \e\,
  (F^1-F^2+i\bar\tau(F^1+F^2))_{\n\r\s}\quad ,
   \\
\delta\ell &=& ie^{-\phi}(\bar\e_C\l - \bar\e\l_C)\quad ,
\\
\delta\phi &=& \bar\e_C \l+\bar\e\l_C\quad .
\end{eqnarray}
For the higher-rank form fields we present only the transformations to the fermions,
because the contributions containing explicit gauge fields are unchanged
by the gauge fixing and redefinitions and can be read off from
(\ref{res-A6}-\ref{res-A10-4}). We find:
\begin{eqnarray}
\d A^1{}_{(6)} &=& 6e^{\phi/2}\big\{
     (1-i\tau)(\bar\e_C\g_{(5)}\psi + \tfrac{i}{12}\bar\e\g_{(6)}\l) \nonumber\\  &&\quad
  -(1-i\bar\tau)(\bar\e\g_{(5)}\psi + \tfrac{i}{12}\bar\e_C\g_{(6)}\l_C) \big\} + \ldots \quad , \label{EdA6-1} \\
\d A^2{}_{(6)} &=& 6e^{\phi/2}\big\{
     (1+i\tau)(\bar\e_C\g_{(5)}\psi + \tfrac{i}{12}\bar\e\g_{(6)}\l) \nonumber\\
    &&\quad -(1+i\bar\tau)(\bar\e\g_{(5)}\psi_{C}+\tfrac{i}{12}\bar\e_C\g_{(6)}\l_C) \big\} + \ldots \quad , \label{EdA6-2} \\
\d A^{11}{}_{(8)} &=& 2 e^{\phi}\big\{
   (1-i\tau)(1-i\bar\tau)(\bar\e\g_{(7)}\psi - \bar\e_C\g_{(7)}\psi_{C}) \nonumber\\
   &&\quad    +\tfrac{i}{8}((1-i\tau)^2\bar\e_C\g_{(8)}\l -
    (1-i\bar\tau)^2\bar\e\g_{(8)}\l_C )\big\} + \ldots\quad , \label{EdA8-11} \\
\d A^{22}{}_{(8)} &=& 2 e^{\phi}\big\{
  (1+i\tau)(1+i\bar\tau)(\bar\e\g_{(7)}\psi -
                         \bar\e_C\g_{(7)}\psi_{C}) \nonumber\\
      &&\quad    +\tfrac{i}{8}((1+i\tau)^2\bar\e_C\g_{(8)}\l -
               (1+i\bar\tau)^2\bar\e\g_{(8)}\l_C)
                        \big\} + \ldots\quad , \label{EdA8-22}\\
\d A^{12}{}_{(8)} &=&  e^{\phi}\big\{ ((1-i\tau)(1+i\bar\tau)
  +(1+i\tau)(1-i\bar\tau))(\bar\e\g_{(7)}\psi -
                         \bar\e_C\g_{(7)}\psi_{C}) \nonumber\\
     && +\tfrac{i}{4}((1-i\tau)(1+i\tau)\bar\e_C\g_{(8)}\l -
               (1-i\bar\tau)(1+i\bar\tau)\bar\e\g_{(8)}\l_C)
                        \big\} + \ldots \, , \label{EdA8-12}
\end{eqnarray}
\begin{eqnarray}
\d A^{1}{}_{(10)} &=&
   10 i e^{\phi/2}\big\{
    (1-i\bar\tau)(\bar\e\g_{(9)}\psi_{C} -
                \tfrac{i}{20}\bar\e_C\g_{(10)}\l_C)
\nonumber\\
&&\quad
   +(1-i\tau)(\bar\e_C\g_{(9)}\psi
       - \tfrac{i}{20}\bar\e\g_{(10)}\l)
                        \big\} \quad ,
\label{EdA10-1}
\\
  \d A^{2}{}_{(10)} &=&
   10 i e^{\phi/2}\big\{
    (1+i\bar\tau)(\bar\e\g_{(9)}\psi_{C} -
                \tfrac{i}{20}\bar\e_C\g_{(10)}\l_C)
\nonumber\\
&&\quad
   +(1+i\tau)(\bar\e_C\g_{(9)}\psi
       - \tfrac{i}{20}\bar\e\g_{(10)}\l)
                        \big\} \quad ,
\label{EdA10-2}
\\
  \d A^{111}_{(10)} &=&
 \tfrac{5}{6}e^{3\phi/2}
 (1-i\tau)(1-i\bar\tau)\big\{
   (1-i\bar\tau)(\bar\e\g_{(9)}\psi_{C} +
                \tfrac{3i}{20}\bar\e_C\g_{(10)}\l_C)
\nonumber\\
&&  -(1-i\tau)(\bar\e_C\g_{(9)}\psi +
                \tfrac{3i}{20}\bar\e\g_{(10)}\l) \big\}
  +\ldots \quad ,
\label{EdA10-111}
\\
\d A^{222}_{(10)} &=&
 \tfrac{5}{6}e^{3\phi/2}
 (1+i\tau)(1+i\bar\tau)\big\{
   (1+i\bar\tau)(\bar\e\g_{(9)}\psi_{C} +
                \tfrac{3i}{20}\bar\e_C\g_{(10)}\l_C)
\nonumber\\
&&  -(1+i\tau)(\bar\e_C\g_{(9)}\psi +
                \tfrac{3i}{20}\bar\e\g_{(10)}\l) \big\}
  +\ldots \quad ,
\label{EdA10-222}
\\
\d A^{112}_{(10)} &=&
 \tfrac{5}{18}e^{3\phi/2}
 \big\{ ((1-i\bar\tau)^2(1+i\tau)+2(1-i\tau)(1-i\bar\tau)(1+i\bar\tau))
       \times
\nonumber\\
&&\quad \times    (\bar\e\g_{(9)}\psi_{C} +
                \tfrac{3i}{20}\bar\e_C\g_{(10)}\l_C)
\nonumber\\
&&  -((1-i\tau)^2(1+i\bar\tau)+2(1-i\tau)(1+i\tau)(1-i\bar\tau))
  \times
\nonumber\\
&&\quad\times
       (\bar\e_C\g_{(9)}\psi +
                \tfrac{3i}{20}\bar\e\g_{(10)}\l)
     \big\} + \ldots \quad ,
\label{EdA10-112}
\\
\d A^{221}_{(10)} &=&
 \tfrac{5}{18}e^{3\phi/2}
 \big\{ ((1+i\bar\tau)^2(1-i\tau)+2(1+i\tau)(1+i\bar\tau)(1-i\bar\tau))
       \times
\nonumber\\
&&\quad\times
     (\bar\e\g_{(9)}\psi_{C} +
                \tfrac{3i}{20}\bar\e_C\g_{(10)}\l_C)
\nonumber\\
&&  -((1+i\tau)^2(1-i\bar\tau)+2(1+i\tau)(1-i\tau)(1+i\bar\tau))
    \times
\nonumber\\
&&\quad\times    (\bar\e_C\g_{(9)}\psi +
                \tfrac{3i}{20}\bar\e\g_{(10)}\l)
     \big\} + \ldots \quad ,
\label{EdA10-221}
\end{eqnarray}
Here the dots stand for the gauge field terms given in (\ref{res-A6}-\ref{res-A10-4}).
So in formulas (\ref{Evielbein}) to (\ref{EdA10-221}) we have collected the complete set of Einstein frame
supersymmetry transformations in the real formulation.\\
Let us now review the transformations under $SU(1,1)$ and $SL(2,\mathbb{R})$ transformations.
Consider an $SU(1,1)$ transformation
\begin{equation}
    U = \left(\begin{array}{cc} \a & \b \\
                               \bar\b & \bar\a
              \end{array}\right)\,, \quad \a\bar\a - \b\bar\b=1\quad .
\end{equation}
The field $\tau$ transforms under the corresponding  $SL(2,\mathbb{R})$ transformation as
\begin{eqnarray}
   \tau &\to& \frac{a\t+b}{c\t+d}\,,\qquad
  \delta\t \to \frac{\delta\t}{(c\t+d)^2}\,,\quad  ad-bc =1\quad,
\nonumber\\
                                     a&=&{\rm Re}(\a-\b)\,,\
                                       d={\rm Re}(\a+\b)\,,\
                                       b=-{\rm Im}(\a+\b)\,,\
                                       c={\rm Im}(\a-\b)\quad \label{sl2rsu11}.
\end{eqnarray}
The redefinition (\ref{redeffermions}) modifies the behavior under $SU(1,1)$ transformations.
The compensating $U(1)$ transformation on a field $\chi$ of charge $q$ (\ref{compU1})
is now changed to
\begin{equation}
     \chi\to e^{iq\xi}\chi\,,\quad {\rm with}\quad
    e^{2i\xi} = \frac{c\t+d}{c\bar\t+d}\quad .
\end{equation}
For the dilaton one finds
\begin{equation}
   e^{\phi}\to e^\phi\,(c\t+d)(c\bar\t+d)\quad .
\end{equation}
One easily verifies that, e.g., the supersymmetry variation of $\tau$
\begin{equation}
   \delta\t = - 2ie^{-\phi}\,\bar\e\l_C
\end{equation}
is consistent with these transformations. The bosonic fields with vanishing $U(1)$
charge still transform in the standard way under $SU(1,1)$.\\

We will now bring some order into the collection of higher-rank forms (\ref{EdA6-1})
to (\ref{EdA10-221}) by considering
certain linear combinations of these. We choose the linear combinations
of the $n$-forms such that for a given $n$, each combination has a unique power of
$\tau$ in the fermionic terms of the supersymmetry variation. This
is motivated by the fact that the RR-forms come with a prefactor of $e^{-\phi}$ in the
standard string frame basis, which is proportional to $\t-\bar{\t}$.
Thus we make the following definitions:
\begin{eqnarray}
\tilde{C}_{(2)} &=& A^1_{(2)} - A^2_{(2)} \qquad, \qquad \tilde{B}_{(2)} = A^1_{(2)} + A^2_{(2)}\quad, \quad\\
\tilde{C}_{(4)} &=& A_{(4)} \quad, \quad \\
\tilde{C}_{(6)} &=& A^1_{(6)} + A^2_{(6)} \qquad, \qquad \tilde{B}_{(6)} = A^1_{(6)} - A^2_{(6)}\quad, \quad\\
\tilde{C}_{(8)} &=& A^{11}_{(8)} + A^{22}_{(8)} + 2 A^{12}_{(8)}, \qquad
\tilde{B}_{(8)} = A^{11}_{(8)} + A^{22}_{(8)} - 2 A^{12}_{(8)} \quad, \quad\\
\tilde{D}_{(8)} &=& A^{11}_{(8)} - A^{22}_{(8)}\quad, \quad\\
\tilde{\mathcal{D}}_{(10)} &=& A^1_{(10)} + A^2_{(10)} ,\qquad
\tilde{\mathcal{E}}_{(10)} = A^1_{(10)} - A^2_{(10)}\quad, \quad\\
\tilde{C}_{(10)} &=& A^{111}_{(10)} + A^{222}_{(10)} + 3\left( A^{112}_{(10)} + A^{221}_{(10)}\right)\quad, \quad\\
\tilde{B}_{(10)} &=& A^{111}_{(10)} - A^{222}_{(10)} - 3\left( A^{112}_{(10)} - A^{221}_{(10)}\right)\quad, \quad\\
\tilde{E}_{(10)} &=& A^{111}_{(10)} + A^{222}_{(10)} - \left( A^{112}_{(10)} + A^{221}_{(10)}\right)\quad, \quad\\
\tilde{D}_{(10)} &=& A^{111}_{(10)} - A^{222}_{(10)} + \left( A^{112}_{(10)} - A^{221}_{(10)}\right) .
\end{eqnarray}

A nice property, and partial justification why we refer to some of these linear
combinations as $\tilde{C}_{(n)}$ ( RR fields) and $\tilde{B}_{(n)}$ ( NS-NS fields)
is the way these fields transform into each other under $S$-duality.
The discrete $S$-duality transformation $\tau\to -1/\tau$ corresponds to an
$SL(2,\mathbb{R})$-transformation with $a=d=0$, $b=-c=1$.
The behaviour of the form-fields under $S$-duality is
\begin{eqnarray}
\tilde{C}_{(2)}  &\to& -i \,   \tilde{B}_{(2)} \quad, \quad \tilde{B}_{(2)}  \to -i\, \tilde{C}_{(2)}  \quad,
\nonumber\\
\tilde{C}_{(4)} &\to& \tilde{C}_{(4)} \quad, \quad \nonumber\\
\tilde{C}_{(6)}  &\to& -i \,   \tilde{B}_{(6)} \quad, \quad \tilde{B}_{(6)}  \to -i\, \tilde{C}_{(6)}  \quad,
\nonumber\\
\tilde{C}_{(8)} &\to& -\, \tilde{B}_{(8)} \quad, \quad
\tilde{B}_{(8)}  \to - \, \tilde{C}_{(8)} \quad, \quad
\tilde{D}_{(8)}  \to - \, \tilde{D}_{(8)} \quad,
\nonumber\\
\tilde{\mathcal{D}}_{(10)} &\to& -i\, \tilde{\mathcal{E}}_{(10)}\, \quad, \quad \tilde{\mathcal{E}}_{(10)} \to   -i\, \tilde{\mathcal{D}}_{(10)} \quad,
\nonumber\\
\tilde{C}_{(10)} &\to& i\,  \tilde{B}_{(10)} \quad,\quad \tilde{B}_{(10)} \to i \,  \tilde{C}_{(10)} \quad,
\nonumber\\
\tilde{D}_{(10)}    &\to& i \tilde{E}_{(10)}
\quad,\quad  \, \tilde{E}_{(10)}   \to i\,\tilde{D}_{(10)} .
\label{Sdualitytable}
\end{eqnarray}
We see that applying $S$-duality twice gives +1 on $\tau$ and on the four- and eight-forms,
but $-1$ on the two-, six- and ten-forms. That this is indeed right, and that the $S$-duality
transformation is not its own inverse can be seen easily from translating back to
the $SU(1,1)$ notation via (\ref{sl2rsu11}), in which the $S$-duality transformation matrix
is given by
\begin{equation}
U = \left(\begin{array}{cc} -i & 0 \\
                               0 & i
              \end{array}\right)\,
\end{equation}
so that $U^2$ gives a minus on forms with an odd number of SU(1,1)-indices.\\
Now we are ready to transform to string frame. The basic transformation
is $e_{(E)\,\m}{}^a = e^{-\phi/4} e_{(S)\,\m}{}^a$. We choose to write the
variation of the zehnbein in standard form, which requires a modification
of supersymmetry with a $\lambda$-dependent local Lorentz transformation (which
we see only on the zehnbein), and a redefinition:
\begin{eqnarray}
  \e' &=& e^{\phi/8}\,\e\quad ,
\\
  \l' &=& e^{-\phi/8}\,\lambda \quad ,
\\
  \psi'_\m &=& e^{\phi/8}\psi_\m - \tfrac{i}{4}\g'_\m\lambda'_C\quad ,
\\
  \g'_\m  &=& e^{\phi/4}\g_\m \quad .
\end{eqnarray}
Again we start with the basic supergravity multiplet and then discuss
the high-rank forms. The transformation rules are simplified by
writing the complex fermions as real doublets, i.e.
$\e\to (\e_1,\  \e_2)$, where $\e_i$ are real Majorana-Weyl fermions.
This gives rise to the appearance of Pauli matrices
$\s_0=1,\s_1,i\s_2,\s_3$ in the contractions
between such doublets, generically:
\begin{eqnarray}
  &&\bar\e_C\g\chi + \bar\e\g\chi_C \to 2\bar\e\sigma_3\g\chi\,,\quad
  \bar\e_C\g\chi_C + \bar\e\g\chi \to 2\bar\e \g\chi\quad ,
\\
  &&\bar\e_C\g\chi - \bar\e\g\chi_C \to 2i\bar\e\sigma_1\g\chi\,,\quad
  \bar\e_C\g\chi_C - \bar\e\g\chi \to -2i\bar\e(i\sigma_2)\g\chi\quad .
\end{eqnarray}
In addition we redefine $\lambda\to \lambda_C$, or, equivalently, in the real
notation
\begin{equation}
   \lambda\to \s_3\lambda\quad .
\end{equation}
We drop all primes in the string frame transformation
rules:
\begin{eqnarray}
\d e_\m{}^a &=& 2i\bar\e\g^a\psi_\m \label{sfvielbein}\\
\d \psi_\m &=&  D_\m \e
       + \tfrac{1}{8}e^{\phi}\g^\n\partial_\n \ell\g_\m \,(i\sigma_2)\,\e
      -\tfrac{1}{16} \g^{\n\r}\s_3\e F_{+ \m\n\r}
\nonumber
\\
 &&
   +\tfrac{i}{96} e^\phi \g^{\n\r\s} \g_\m \,\s_1\,\e (F_- + i\ell F_+ )_{\n\r\s}
\nonumber\\
&& -\tfrac{1}{480} e^\phi  \g^{\m_1\ldots\m_5}\gamma_\mu\,(i\sigma_2)\,\e
      F_{\m_1\ldots\m_5}
\quad ,
\\
\delta \tb_{\m\n} &=&
  8i \,\bar\e\,\s_3\,\g_{[\m}\psi_{\n]} \quad ,
\\
\delta \tc_{\m\n} &=&
  -8 e^{-\phi} \,\bar\e\,\s_1\, \g_{[\m}(\psi_{\n]}+\tfrac{i}{2}\g_{\n]}\l)
 -i\ell\,\d \tb_{\m\n}
\quad ,
\\
\d \tc_{\m\n\r\s}&=& 2i e^{-\phi}
   \,\bar\e\,(i\s_2)\,\g_{[\m\n\r}
   (\psi_{\s]}+\tfrac{i}{4}\g_{\s]}\l)
\nonumber\\
&&   -\tfrac{3i}{16}\{ \tc \d \tb-\tb\d\tc\}_{\m\n\r\s}
    \quad ,
\\
\d \lambda &=& \tfrac{i}{2}\g^\m\partial_\m\phi\,\e
     -\tfrac{i}{48}\g^{\n\r\s}\,\s_3\,\e F_{+ \n\r\s}
   -\tfrac{i}{2}e^\phi \g^\m\partial_\m\ell\,(i\sigma_2)\,\e
\nonumber\\
&&
   +\tfrac{1}{48}e^\phi\g^{\n\r\s}\,\s_1\,\e (F_- +i\ell F_+ )_{\n\r\s}
  \quad ,
\\
\delta\ell &=& 2\,e^{-\phi}\bar\e\,(i\s_2)\,\l\quad ,
\\
\delta\phi &=& 2\,\bar\e\l\quad \label{sfphi}
\end{eqnarray}
where we have defined
\be
F_+ = F^1 + F^2, \qquad F_- = F^1 - F^2 .
\ee
For the higher form fields we find:
\begin{eqnarray}
\d \tilde{C}_{(6)} &=& 24i\, e^{-\phi}\,
    \bar\e\,\sigma_1\g_{\,(5)}(\psi +\tfrac{i}{6}\g_{(1)}\l)
    + 40 \tc_{(4)} \d \tilde{B}_{(2)} \nonumber\\
    &&-20 \d \tc_{(4)} \tilde{B}_{(2)} -\tfrac{15i}{4} \tilde{B}_{(2)}
    \left( \tc_{(2)}\d\tb_{(2)} - \tb_{(2)}\d\tc_{(2)} \right) \quad , \label{SFres-A61p2} \\
\d \tilde{B}_{(6)} &=& 24\, e^{-\phi}\, \big\{
   \ell\,\bar\e\,\sigma_1 \g_{\,(5)}(\psi + \tfrac{i}{6}\g_{(1)}\l)
   +\,e^{-\phi}\,\bar\e\,\sigma_3 \g_{\,(5)}(\psi +\tfrac{i}{3}\g_{(1)}\l) \big\} \nonumber\\
   && + 40 \tc_{(4)} \d \tilde{C}_{(2)} -20 \d \tc_{(4)} \tilde{C}_{(2)}
-\tfrac{15i}{4} \tilde{C}_{(2)} \left( \tc_{(2)}\d\tb_{(2)} - \tb_{(2)}\d\tc_{(2)} \right)\, , \label{SFres-A61m2}\\
\d \tilde{C}_{(8)} &=& 16i\, e^{-\phi}\,
    \bar\e\,(i\sigma_2)\g_{\,(7)} (\psi +\tfrac{i}{8}\g_{(1)}\l)
   +\tfrac{21}{4} \tilde{C}_{(6)} \d \tilde{B}_{(2)}
   - \tfrac{7}{4}  \tilde{B}_{(2)}\d \tilde{C}_{(6)} \nonumber \\
    && -35 \tilde{B}_{(2)} \tilde{B}_{(2)} \d \tc_{(4)}
    + 70 \tc_{(4)} \tilde{B}_{(2)} \d \tilde{B}_{(2)]}  \nonumber\\
  && -  \tfrac{105i}{16} \tilde{B}_{(2)} \tilde{B}_{(2)} \left( \tc_{(2)}\d\tb_{(2)} - \tb_{(2)}\d\tc_{(2)} \right)
   \quad , \label{SFres-A8pp} \\
\d \tilde{B}_{(8)} &=& -16i\, e^{-\phi}\,\{ \ell^2\,\bar\e\,(i\sigma_2)\g_{\,(7)}
        (\psi + \tfrac{i}{8}\g_{(1)}\l) \nonumber\\
     &&    +\tfrac{i}{4}\ell e^{-\phi}\,\bar\e \g_{(8)}\lambda
    +e^{-2\phi}\,\bar\e\,(i\sigma_2) \g_{\,(7)} (\psi +\tfrac{3i}{8}\g_{(1)}\l) \} \nonumber\\
   && + \tfrac{21}{4} \tilde{B}_{(6)} \d \tilde{C}_{(2)}
    - \tfrac{7}{4}  \tilde{C}_{(2)} \d \tilde{B}_{(6)}
     -35 \tilde{C}_{(2)} \tilde{C}_{(2)} \d \tc_{(4)}  \nonumber\\
    && + 70 \tc_{(4)} \tilde{C}_{(2)} \d \tilde{C}_{(2)}
    - \tfrac{105i}{16} \tilde{C}_{(2)} \tilde{C}_{(2)} \left( \tc_{(2)}\d\tb_{(2)} - \tb_{(2)}\d\tc_{(2)} \right)
   \quad , \label{SFres-A8pm}\\
\d \tilde{D}_{(8)} &=& 16 \ell e^{-\phi}\, \bar\e\,(i\sigma_2)\g_{\,(7)}
            (\psi + \tfrac{i}{8}\g_{(1)}\l) + 2ie^{-2\phi} \bar\e\g_{(8)}\l  \nonumber\\
     && +\tfrac{21}{8}\{ \tilde{C}_{(6)}
                     \d \tilde{C}_{(2)} + \tilde{B}_{(6)} \d \tilde{B}_{(2)} \} \nonumber\\
     &&- \tfrac{7}{8} \{ \tilde{B}_{(2)}\d \tilde{B}_{(6)}
     + \tilde{C}_{(2)}\d \tilde{C}_{(6)} \}
   -35 \tilde{B}_{(2)} \tilde{C}_{(2)} \d \tc_{(4)}  \nonumber\\
   && + 35 \tc_{(4)}\{\tilde{B}_{(2)} \d \tilde{C}_{(2)} +\tilde{C}_{(2)} \d \tilde{B}_{(2)} \} \nonumber\\
  && -  \tfrac{105i}{16} \tilde{B}_{(2)} \tilde{C}_{(2)}\left( \tc_{(2)}\d\tb_{(2)} - \tb_{(2)}\d\tc_{(2)} \right) \quad,
\label{SFres-A8m}
\end{eqnarray}
\begin{eqnarray}
\d\tilde{\mathcal{D}}_{(10)} &=&
   40i\, e^{-2\phi}\, \bar\e\,\sigma_3\g_{\,(9)}
            (\psi + \tfrac{i}{5}\g_{(1)}\l) \quad , \label{SFres-A101p2} \\
\d \tilde{\mathcal{E}}_{(10)} &=& 40\, e^{-2\phi}\,
    \big\{ \ell\, \bar\e\,\sigma_3\g_{\,(9)} (\psi +\tfrac{i}{5}\g_{(1)}\l)
    - e^{-\phi}\,\bar\e\,\sigma_1\g_{\,(9)} (\psi +\tfrac{3i}{10}\g_{(1)}\l)\big\}
\quad, \label{SFres-A101m2}\\
\d\tilde{C}_{(10)} &=& - \tfrac{40}{3} i e^{-\f} \bar{\e}\sigma_1\g_{(9)}\left(\psi+
 \tfrac{i}{10} \g_{(1)} \l \right) \\
 &&-12 \tilde{C}_{(8)} \d \tilde{B}_{(2)}
   + 3 \d \tilde{C}_{(8)} \tilde{B}_{(2)}
-\tfrac{63}{4} \tc_{(6)} \tb_{(2)} \d \tb_{(2)}
   +\tfrac{21}{4} \d \tc_{(6)} \tb_{(2)} \tb_{(2)}\nonumber \\
   &&-210 \tc_{(4)}\tb_{(2)} \tb_{(2)} \d \tb_{(2)}
   + 105 \d \tc_{(4)}\tb_{(2)} \tb_{(2)} \tb_{(2)}\nonumber\\
   &&+ \tfrac{315}{16} i \tb_{(2)} \tb_{(2)} \tb_{(2)}
   \left(\tc_{(2)} \d\tb_{(2)} - \tb_{(2)} \d\tc_{(2)}\right) \quad ,\\
\d\tilde{B}_{(10)} &=&
  +\tfrac{40}{3} e^{-4\f} \left(1 + e^{2\f}\ell^2\right) \bar{\e}\sigma_3\g_{(9)}\left(\psi+
  \tfrac{2i}{5} \g_{(1)} \l \right)\nonumber\\
  &&+\tfrac{40}{3} \ell e^{-\f} \left(\ell^2 + e^{-2\f}\right) \bar{\e}\sigma_1\g_{(9)}\left(\psi+
  \tfrac{i}{10} \g_{(1)} \l \right)\nonumber\\
 && -12 \tb_{(8)} \d \tc_{(2)}
   + 3 \d \tb_{(8)} \tc_{(2)}
-\tfrac{63}{4} \tb_{(6)} \tc_{(2)} \d \tc_{(2)}
   +\tfrac{21}{4} \d \tb_{(6)} \tc_{(2)} \tc_{(2)}\nonumber \\
   && -210 \tc_{(4)}\tc_{(2)} \tc_{(2)} \d \tc_{(2)}
   + 105 \d \tc_{(4)}\tc_{(2)} \tc_{(2)} \tc_{(2)}\nonumber\\
   &&+ \tfrac{315}{16} i \tc_{(2)} \tc_{(2)} \tc_{(2)}
   \left(\tc_{(2)} \d\tb_{(2)} - \tb_{(2)} \d\tc_{(2)}\right) \quad ,\\
\d\tilde{D}_{(10)} &=&
  -\tfrac{40}{9} e^{-2\f} \bar{\e}\sigma_3\g_{(9)}\left(\psi+
  \tfrac{2i}{5} \g_{(1)} \l \right)
 -\tfrac{40}{3} \ell e^{-\f} \bar{\e}\sigma_1\g_{(9)}\left(\psi+
  \tfrac{i}{10} \g_{(1)} \l \right)\nonumber\\  
&&+2 \tilde{B}_{(2)} \d \tilde{D}_{(8)} +   \tilde{C}_{(2)} \d \tilde{C}_{(8)}
               -8 \tilde{D}_{(8)} \d \tilde{B}_{(2)} - 4 \tilde{C}_{(8)} \d \tilde{C}_{(2)}
\nonumber\\
&& + \tfrac{7}{4}\,\big(\tilde{B}_{(2)}^2 \d \tilde{B}_{(6)} + 2 \tilde{B}_{(2)}\tilde{C}_{(2)}\d \tilde{C}_{(6)}\big)
\nonumber\\
&& -\tfrac{21}{4}\,\big( \tilde{B}_{(2)}\tilde{C}_{(6)}\d \tilde{C}_{(2)} + \tilde{C}_{(2)}\tilde{C}_{(6)}\d \tilde{B}_{(2)}
               + \tilde{B}_{(2)}\tilde{B}_{(6)}\d \tilde{B}_{(2)}\big)              
\nonumber\\
&&  - 70\,\big(\tilde{B}_{(2)}^2 \tilde{C}_{(4)} \d \tilde{C}_{(2)} 
       + 2 \tilde{B}_{(2)}\tilde{C}_{(2)}\tilde{C}_{(4)} \d \tilde{B}_{(2)}
     - \tfrac{3}{2}\tilde{B}_{(2)}^2 \tilde{C}_{(2)} \d \tilde{C}_{(4)}\big)
\nonumber\\ 
&& + \tfrac{315}{16}i\,\big(\tilde{B}_{(2)}^2\tilde{C}_{(2)}^2\d \tilde{B}_{(2)}
                     -\tilde{B}_{(2)}^3\tilde{C}_{(2)}  \d \tilde{C}_{(2)}\big)\,,
\nonumber\\    
\d\tilde{E}_{(10)} &=&
 +\tfrac{40}{3} i e^{-3\f} \left(\tfrac{1}{3} + \ell^2 e^{2\f} \right) \bar{\e}\sigma_1\g_{(9)}\left(\psi+
  \tfrac{i}{10} \g_{(1)} \l \right) \nonumber\\
 &&+\tfrac{80}{9} i \ell e^{-2\f}  \bar{\e}\sigma_3\g_{(9)}\left(\psi+
  \tfrac{2i}{5} \g_{(1)} \l \right)\nonumber\\
&&+ 2 \tilde{C}_{(2)} \d \tilde{D}_{(8)} +   \tilde{B}_{(2)} \d \tilde{B}_{(8)}
               -8 \tilde{D}_{(8)} \d \tilde{C}_{(2)} - 4 \tilde{B}_{(8)} \d \tilde{B}_{(2)} 
\nonumber\\ 
&& + \tfrac{7}{4}\,\big(\tilde{C}_{(2)}^2 \d \tilde{C}_{(6)} + 2 \tilde{B}_{(2)}\tilde{C}_{(2)} \d \tilde{B}_{(6)}\big)
\nonumber\\
&&  -\tfrac{21}{4}\,\big( \tilde{B}_{(2)}\tilde{B}_{(6)}\d \tilde{C}_{(2)} + \tilde{C}_{(2)}\tilde{C}_{(6)}\d \tilde{C}_{(2)}
               + \tilde{C}_{(2)}\tilde{B}_{(6)}\d \tilde{B}_{(2)}\big)  
\nonumber\\
&&  - 70\,\big(\tilde{C}_{(2)}^2 \tilde{C}_{(4)} \d \tilde{B}_{(2)} 
         + 2 \tilde{B}_{(2)}\tilde{C}_{(2)}\tilde{C}_{(4)}  \d \tilde{C}_{(2)}
     - \tfrac{3}{2}\tilde{B}_{(2)} \tilde{C}_{(2)}^2 \d \tilde{C}_{(4)}\big)   
\nonumber\\
&&  + \tfrac{315}{16}i\,\big(\tilde{B}_{(2)}\tilde{C}_{(2)}^3 \d \tilde{B}_{(2)}
                     -\tilde{B}_{(2)}^2\tilde{C}_{(2)}^2\d \tilde{C}_{(2)}\big)\, .
\end{eqnarray}

We will now introduce the standard RR and NS-NS fields, and extend this to the
higher rank forms.
For this we define:
\begin{eqnarray}
   B_{(2)} &=& \tfrac{1}{2}\tb_{(2)}\quad ,
\\
   C_{(0)} &=& -\tfrac{1}{2}\ell\quad,
\\
   C_{(2)} &=& -\tfrac{i}{4}\tc_{(2)}\quad ,
\\
  C_{(4)} &=& 2 \tc_{(4)} + 3 C_{(2)}B_{(2)}\quad,
\end{eqnarray}
for which we define curvatures
\begin{eqnarray}
  H_{(3)}    &=& 3\partial B_{(2)}\quad ,
\\
  G_{(2n-1)} &=& (2n-1)\{\partial C_{(2n-2)} - \tfrac{1}{2}(2n-2)(2n-3)
         C_{(2n-4)}\partial B_{(2)} \}\quad .
\label{curvatures}
\end{eqnarray}
In (\ref{curvatures}) $n$ takes on the values $n=1,2,3$, but this will be extended
to $n\le 6$ below.
The corresponding bosonic gauge transformations are
\begin{eqnarray}
   \d B_{(2)} &=& \partial \Sigma\quad ,
\\
   \d C_{(2n-2)} &=& \partial \Lambda_{(2n-3)}
         + \tfrac{1}{2}(2n-2)(2n-3)\Lambda_{(2n-5)}\partial B_{(2)}\quad .
\end{eqnarray}
We now rewrite the supergravity multiplet in these variables:
\begin{eqnarray}
  \d e_\m{}^a &=& 2i\bar\e\g^a\psi_\m
\\
\d \psi_\m &=&  D_\m \e -\tfrac{1}{8} \g^{\n\r}\,\s_3\,\e H_{\m\n\r}
    - \tfrac{1}{4}e^{\phi}(\g\cdot G_{(1)}) \g_\m \,(i\sigma_2)\,\e
\nonumber
\\
 &&
   - \tfrac{1}{24} e^\phi (\g\cdot G_{(3)}) \g_\m \,\s_1\,\e
   -\tfrac{1}{960} e^\phi (\g\cdot G_{(5)}) \g_\m\,(i\s_2)\,\e
\quad ,
\\
\delta B_{\m\n} &=&
     4i \,\bar\e\,\s_3\,\g_{[\m}\psi_{\n]} \quad ,
\\
\delta C_{(0)} &=& - e^{-\phi}\bar\e\,(i\s_2)\,\l\quad ,
\\
\delta C_{\m\n} &=&
    2i  e^{-\phi} \,\bar\e\,\s_1\, \g_{[\m}(\psi_{\n]}+\tfrac{i}{2}\g_{\n]}\l)
   + C_{(0)}\delta B_{\m\n} \quad ,
\\
\d C_{\m\n\r\s}&=&
    4i e^{-\phi}\,\bar\e\,(i\s_2)\,\g_{[\m\n\r}(\psi_{\s]}+\tfrac{i}{4}\g_{\s]}\l)
 + 6 C_{[\m\n} \delta B_{\r\s]}   \quad ,
\\
\d \lambda &=& \tfrac{i}{2}\g^\m\partial_\m\phi\,\e
  -\tfrac{i}{24}(\g\cdot H_{(3)}) \,\s_3\,\e
\nonumber\\
&&   + {i} e^\phi (\g\cdot G_{(1)}) \,(i\sigma_2)\,\e
   +\tfrac{i}{12}e^{\phi} (\g\cdot G_{(3)}) \,\s_1\,\e \quad ,
\\
\delta\phi &=& 2\,\bar\e\l\quad .
\end{eqnarray}
The supersymmetry transformations of the RR fields $C$ can be summarized
as ($n=1,2,3,\ {\cal P}_n=i\s_2$ for $n$ even, ${\cal P}_n=\s_1$ for $n$ odd):
\begin{eqnarray}
  \delta C_{(2n-2)} &=& (2n-2)i e^{-\phi}
    \bar\e \,{\cal P}_{n}\, \g_{(2n-3)} (\psi_{(1)}+\tfrac{i}{2n-2}\g_{(1)}\l)
\nonumber\\
&&\qquad   + \tfrac{1}{2}(2n-2)(2n-3) C_{(2n-4)}\delta B_{(2)}\quad .
\label{RRharmonica}
\end{eqnarray}

We will now extend this to the higher-rank forms. We define the following
RR fields:
\begin{eqnarray}
   C_{(6)} &=& \tfrac{1}{4}\tc_{(6)} + 5 C_{(4)}B_{(2)}\quad ,
\\
   C_{(8)} &=& \tfrac{1}{2}\tc_{(8)} + 7C_{(6)}B_{(2)}\quad ,
\\
   C_{(10)}&=& -\tfrac{3}{4}\tc_{(10)} + 9C_{(8)}B_{(2)}\quad .
\end{eqnarray}
These combinations transform precisely as (\ref{RRharmonica}). We have
therefore identified the tower of RR forms, in the same form as in
\cite{BdRJO}. The $S$-dual of $C_{(10)}$ is however not the field
$B_{(10)}$ given in \cite{BdRJO}. It turns out that $B_{(10)}$ corresponds
precisely to our $\tilde{\mathcal{D}}_{(10)}$.

The $S$-duals of the $C_{(2n-2)}$ should  form a tower of NS-NS forms.
If one defines that under $S$-duality
\begin{eqnarray}
  &&C_{(2)}\to iS_{(2)}\,,
\nonumber\\
  &&C_{(4)}\to  S_{(4)}\,,
\nonumber\\
 &&C_{(6)}\to -iS_{(6)}\,,
\nonumber\\
 && C_{(8)}\to - S_{(8)}\,,
\nonumber\\
 &&C_{(10)}\to  iS_{(10)}\,,
\end{eqnarray}
then we find
\begin{eqnarray}
   S_{(2)} &=& \tfrac{i}{4}\tilde{B}_{(2)}\,,\\
   S_{(4)} &=& 2 \tc_{(4)} +6i C_{(2)} S_{(2)}\,,\\
   S_{(6)} &=& \tfrac{1}{4} \tb_{(6)} + 10 i C_{(2)} S_{(4)} \,,\\
   S_{(8)} &=& \tfrac{1}{2} \tb_{(8)} + 14 i C_{(2)} S_{(6)}\,,\\
   S_{(10)} &=& -\tfrac{3}{4} \tb_{(10)} + 18 i C_{(2)} S_{(8)}\,.
\end{eqnarray}
For the case $\ell=0$ the supersymmetry variations for $S_{(n)}$ are
then described by
\begin{eqnarray}
\d S_{(2n-2)} &=& (-i)^n (2n-2) e^{- (n-2) \f} \bar{\e} {\cal{P}}_n \g_{(2n-3)} \left(\psi +
\tfrac{n-2}{2n-2}\g_{(1)}\l \right) \\
&& + i (2n-2) (2n-3) S_{(2n-4)}\d C_{(2)}
\end{eqnarray}
where $S_{(0)}=0$ and ${\cal{P}}_n = \s^3$ for $n$ even and ${\cal{P}}_n = i \s^2$ for $n$ odd.

\section{Summary and Discussion}

In this work we showed that the standard formulation of IIB
supergravity can be extended to include a doublet and a quadruplet
of ten-form potentials. We argued that no other independent
ten-forms can be added to the algebra. We have been using a
``$SU(1,1)$-democratic'' formulation, in which {\it all} forms are
described together with their magnetic duals. Furthermore, all forms
transform in a given representation under the duality group
$SL(2,\mathbb{R})$. The previously known RR-ten-form potential
$C_{(10)}$ is contained in the quadruplet. The other previously
known ten-form (named $B_{(10)}$ in \cite{BdRJO}) is in the doublet
and hence not $S$-dual to $C_{(10)}$ ~\cite{fabio}. 

We have shown that all ten-form potentials have a leading term
\be
\d X_{(10)} \sim e^{n \phi} \bar{\e} \g_{(9)} \psi \,\,\,\, \text{at} \,\, l=0
\ee
in their supersymmetry transformation in string frame where $X_{(10)}$ represents a
generic ten-form potential.

\begin{table}[h]
\begin{center}
\renewcommand{\arraystretch}{1.5}
\begin{tabular}{|c|c|c|c|}
\hline
potential in quadruplet & potential in doublet & associated brane & tension\\
\hline
$C_{(10)}$ & & D9-brane & $g_S^{-1}$\\
$D_{(10)}$ & $\mathcal{D}_{(10)}$& solitonic brane & $g_S^{-2}$\\ 
$E_{(10)}$ & $\mathcal{E}_{(10)}$& exotic & $g_S^{-3}$\\ 
$B_{(10)}$ & & exotic & $g_S^{-4}$\\
\hline
\end{tabular}
\renewcommand{\arraystretch}{1.0}
\end{center}
\caption[Table]{Ten-form potentials in string frame, the corresponding branes and their
tension in terms of the string coupling $g_S$.}
\label{couplingtable}
\end{table}

Such ten-form potentials naturally occur as the leading contribution in Wess-Zumino
terms for space-time filling branes with tension $g_S^n$. The resulting
branes can be found in table \ref{couplingtable}. These branes and their
relevance for theories with sixteen supercharges will be discussed in some
detail in a forthcoming paper \cite{BDKOR}.

It would be
interesting to see how these findings are compatible with the known
$S$-duality relations between the Heterotic and Type I superstrings.
It is well-known that the (Nambu-Goto part of the) tree-level action
of the Type I (Heterotic) superstring scales with $g_S^{-1}$
($g_S^{-4}$) \cite{Bergshoeff98}. The interpretation of the
$g_S^{-4}$ term at the Heterotic side is not clear. However, the
results presented in Table \ref{truncationtable}, Appendix
\ref{truncations}, and the $S$-duality assignments of the ten-forms
(\ref{Sdualitytable}) open up the possibility to extend this,
consistent with $S$-duality, to the scaling behaviour  $g_S^{-1} +
g_S^{-3}$ for the Type I superstring and  $g_S^{-2} + g_S^{-4}$ for
the Heterotic superstring such that the Nambu-Goto term at the
Heterotic side contains the more conventional $g_S^{-2}$ behaviour.

Work on the relation of string- and M-theory with the Kac-Moody
algebras $E_{11}$ \cite{west, schnakenburgwest} and $E_{10}$
\cite{damourhenneauxnicolai,damournicolai} 
has an interesting connection with
our results. In \cite{kleinschmidtnicolai}
it was pointed out that $E_{10}$ and $E_{11}$ give rise to different
IIB ten-form potentials. In particular,
$E_{10}$ does not give rise to ten-forms, whereas  $E_{11}$
supports a doublet and
a quadruplet of ten-forms \cite{kleinschmidtwestschnakenburg}.
The latter is in agreement with our results.

It will be worthwile to derive the superspace formulation of
our results. Note that, although ten-form potentials have identically
zero field-strengths, this is not true for the ten-form
superpotentials. It would be interesting to calculate the eleven-form
curvatures in flat superspace and to see to which kind of
Wess-Zumino terms they give rise to. This is the first step towards
the construction of a kappa-symmetric Green-Schwarz action for {\it
all} 9-branes.

\section{Acknowledgements}

We thank Axel Kleinschmidt, Hermann Nicolai and Tomas
Ort\'\i n for useful remarks.\\
E.B., S.K. and M. de R. are supported
by the  European Commission FP6 program
MRTN-CT-2004-005104 in which E.B., S.K. and M. de R. are
associated to Utrecht university.
S.K. is supported by a Postdoc-fellowship of the German
Academic Exchange Service (DAAD).
F.R. is supported by a European
Commission Marie Curie Postdoctoral Fellowship, Contract
MEIF-CT-2003-500308.
The work of E.B. is partially supported by the Spanish grant
BFM2003-01090.

\begin{appendix}
\section{Conventions}

The Levi-Civita symbol used in this paper is a tensor,
and therefore includes the appropriate powers of $\det\ e$.

Some useful properties of the complex
fermions are:
\begin{eqnarray}
   \psi_\m&=& -\gamma_{11} \psi_\m\,,
\\
   \lambda    &=& \gamma_{11}\lambda\,,
\\
   D_\m\epsilon &=& (\partial_\m +
     \tfrac{1}{4}\omega_\m{}^{ab}\gamma_{ab}
     -\tfrac{i}{2}Q_\m)\epsilon\,,
\\
   (\bar\chi_1\gamma^{\m_1\ldots\m_n}\chi_2)^*
    &=& \bar\chi_2\gamma^{\m_n\ldots\m_1}\chi_1
    = (-1)^n
    \bar\chi_{1C}\gamma^{\m_1\ldots\m_n}\chi_{2C}\,,
\label{pairconj}
\\
   \bar\chi_1\gamma^{\m_1\ldots\m_n}\chi_2 &=&
   (-1)^{n(n+1)/2}
   \bar\chi_{2C}\gamma^{\m_1\ldots\m_n}\chi_{1C}\quad .
\label{pairflip}
\end{eqnarray}
In these equations $\chi_i$ are arbitrary spinors, not necessarily
Majorana or Weyl.\\
For the duality transformations of $\g$-matrices we have:
\begin{equation}
\g^{\m_1 \ldots \m_n} =
- (-1)^{\tfrac{1}{2} n(n-1)} \tfrac{1}{(10-n)!}
\e^{\m_1\ldots\m_{10}} \g_{\m_{n+1}\ldots \m_{10}}\g_{11}
\label{gammaduality}
\end{equation}

The table below gathers the values of the $U(1)$ weights of the different fields.
The zehnbein $e_\m{}^a$ and all form-fields $A_{(2n)}$ have weight zero.
\begin{table}[h]
\begin{center}
\renewcommand{\arraystretch}{1.5}
\begin{tabular}{|l|c||l|c||l|c||l|c|}
\hline
$V_+^\a$  &  1            &$G_{(3)}$&1& $A_{(n)}$&0&& \\
$V_-^\alpha$  &  $-1$     &$G_{(7)}$&1&&&& \\
$\epsilon$    &$\tfrac{1}{2}$  &$\epsilon_C$  &$-\tfrac{1}{2}$& $\bar\epsilon$&$-\tfrac{1}{2}$&$\bar\epsilon_C$&$\tfrac{1}{2}$\\
$\psi_\m$    &$\tfrac{1}{2}$  &$\psi_{C\,\m}$ &$-\tfrac{1}{2}$& $\bar\psi_\m$&$-\tfrac{1}{2}$ &$\bar\psi_{\,C\m}$&$+\tfrac{1}{2}$\\
$\lambda$        &$\tfrac{3}{2}$  &$\lambda_C$      &$-\tfrac{3}{2}$& $\bar\lambda    $&$-\tfrac{3}{2}$ &$\bar\lambda_C    $&$\tfrac{3}{2}$\\
\hline
\end{tabular}
\renewcommand{\arraystretch}{1.0}
\end{center}
\caption[Table]{Table of $U(1)$ weights
\label{U1}}
\end{table}
\vspace{.3truecm}

\section{Truncations\label{truncations}}

We briefly sketch how to apply the heterotic and type I truncations \cite{BdRJO}
to our IIB results and give a list of the fields surviving the truncation.\\
We first express the complex spinor $\e$ in terms of two real spinors
\begin{equation}
\e = \e_1 + i \e_2.
\end{equation}
The heterotic truncation is then given by setting
\begin{equation}
\e = \pm \e_C . \label{trunc}
\end{equation}
We will work with the "+" choice. We also need to make a choice of gauge for the scalars.
We make the same choice as in section \ref{SF}:
\begin{equation}
V_+^2 = V_-^1.
\end{equation}
Plugging (\ref{trunc}) into the SUSY variation of  $\psi$ we find
\begin{equation}
\psi = \psi_C .
\end{equation}
Similarly, we use the SUSY variations of the other fields to find how the truncation acts on all the fields
\begin{eqnarray}
\psi &=& \psi_C, \,\,\,\,\, \l = \l_C \, ,\label{trunc2}\\
V_+^2 &=& V_-^1 , \,\,\,\,\, V_-^2 = V_+^1 \,,\label{hetscalars} \\
A^1_{(2)} &=& A^2_{(2)} \, ,\\
A_{(4)} &=& 0 \, ,\\
A^{1}_{(6)} &=& - A^{2}_{(6)}\, ,\\
A^{11}_{(8)} &=& - A^{22}_{(8)} , \,\,\,\,\, A^{12}_{(8)} = 0 \, ,\\
A^1_{(10)} &=&  A^2_{(10)} \, ,\\
A^{111}_{(10)} &=& - A^{222}_{(10)} , \,\,\,\,\, A^{112}_{(10)} = - A^{122}_{(10)} \, .
\end{eqnarray}
We also observe that the relations for the scalars (\ref{hetscalars}) imply, using the reality
properties of the scalars and (\ref{defz}), that $z = \bar{z}$. This implies that the axion is
eliminated by the truncation, using (\ref{deftau}) and (\ref{deftlf}).\\

The type I truncation is given by setting
\begin{equation}
\e = \pm i \e_C
\end{equation}
where we work with the "+"-choice again. We choose $V_+^2 = V_-^1$ again and find
from the SUSY variations
\begin{eqnarray}
\e &=& i \e_C , \,\,\,\, \psi = i \psi_C , \,\,\,\, \l = -i \l_C \, ,\\
V_+^2 &=& V_-^1 ,\,\,\,\, V_-^2 = V_+^1 \\
A^1_{(2)} &=& - A^2_{(2)} \, ,\\
A_{(4)} &=& 0 \, ,\\
A^{1}_{(6)} &=& A^{2}_{(6)}\, ,\\
A^{11}_{(8)} &=& - A^{22}_{(8)}, \,\,\,\, A^{12}_{(8)} = 0 \, ,\\
A^1_{(10)} &=& - A^2_{(10)}\, ,\\
A^{111}_{(10)} &=& A^{222}_{(10)}, \,\,\,\, A^{112}_{(10)} = A^{122}_{(10)}\, .
\end{eqnarray}
As in the case of the heterotic truncation, the axion is eliminated by the truncation.
We collect the surviving fields of both truncations in table \ref{truncationtable}.

\begin{table}[h]
\begin{center}
\renewcommand{\arraystretch}{1.5}
\begin{tabular}{|l||l|}
\hline
type I truncation & heterotic truncation\\
\hline
$\phi $ & $\phi $\\
$\tilde{C}_{(2)} \sim e^{-\f}$ & $\tilde{B}_{(2)} \sim e^{0\f}$\\
$\tilde{C}_{(6)} \sim e^{-\f}$ & $\tilde{B}_{(6)} \sim e^{-2\f}$ \\
$\tilde{D}_{(8)} \sim e^{-2\f}$ & $\tilde{D}_{(8)} \sim e^{-2\f}$\\
$\tilde{\mathcal{E}}_{(10)} \sim e^{-3\f} $ & $\tilde{\mathcal{D}}_{(10)} \sim e^{-2\f}$\\
$\tilde{C}_{(10)} \sim e^{-\f}$ & $\tilde{B}_{(10)} \sim e^{-4\f}$\\
$\tilde{E}_{(10)} \sim e^{-3\f}$ & $\tilde{D}_{(10)} \sim e^{-2\f}$\\
\hline
\end{tabular}
\renewcommand{\arraystretch}{1.0}
\end{center}
\caption[Table]{ Field contents of the type I and heterotic truncations. After the
tilde we indicate how the field scales with respect to the dilaton. The entries in
every line are $S$-dual to each other (up to a factor).} \label{truncationtable}
\end{table}
\end{appendix}

\end{document}